\documentclass[]{aa}
\usepackage{graphicx}
\usepackage[normalem]{ulem} 
\usepackage{txfonts}
\usepackage{amssymb}
\usepackage{enumerate}
\usepackage{subcaption}

\usepackage[colorlinks=true,allcolors=blue]{hyperref}%
\newcommand{\llnr}[1]{{\bf \color{magenta}{[]}} \color{black}} 

\usepackage{xcolor}
\usepackage{soul}
\definecolor{mydarkgreen}{RGB}{0,100,0}

\usepackage{academicons}
\definecolor{orcidlogocol}{HTML}{A6CE39}

\newcommand{\orcid}[1]{\href{https://orcid.org/#1}{\textcolor[HTML]{A6CE39}{\aiOrcid}}}

\begin{document}

 \title{Toroidal Miller-Turner and Soloviev CME models in EUHFORIA: I. Implementation}

 \author{L. Linan\href{https://orcid.org/0000-0002-4014-1815}{\includegraphics[scale=0.05]{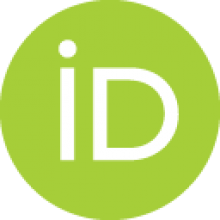}}\inst{1} \and A. Maharana\href{https://orcid.org/0000-0002-4269-056X}{\includegraphics[scale=0.05]{orcid-ID.png}}\inst{1,2} \and S. Poedts\href{https://orcid.org/0000-0002-1743-0651}{\includegraphics[scale=0.05]{orcid-ID.png}}\inst{1,3} \and B. Schmieder
 \href{https://orcid.org/0000-0003-3364-9183}{\includegraphics[scale=0.05]{orcid-ID.png}}\inst{1,4,5} \and R. Keppens\href{https://orcid.org/0000-0003-3544-2733}{\includegraphics[scale=0.05]{orcid-ID.png}} \inst{1}}
 
 \institute{Centre for mathematical Plasma-Astrophysics, Department of Mathematics, KU Leuven, Celestijnenlaan 200B, 3001 Leuven, Belgium \\
 \and Solar–Terrestrial Centre of Excellence—SIDC, Royal Observatory of Belgium, 1180 Brussels, Belgium\\
  \and Institute of Physics, University of Maria Curie-Sk{\l}odowska, ul.\ Radziszewskiego 10, 20-031 Lublin, Poland \\
    \and LESIA, Observatoire de Paris, Université PSL, CNRS, Sorbonne Université, Université de Paris, 5 place Jules Janssen, 92190 Meudon, France\\
  \and University of Glasgow, School of Physics and Astronomy, Glasgow, G128QQ, Scotland}

 \date{Received ; accepted}

 
 \abstract
 {EUHFORIA is a space weather forecasting tool used to predict the time of arrival and geo-effectiveness of coronal mass ejections (CMEs). In this simulation model, magnetic structures evolve in the heliosphere after their injection into the domain at 0.1~AU. The predictions provided by EUHFORIA are directly related to the geometric, thermodynamic, and magnetic properties of the injected CME models.}
 {The aim of this paper is to present the implementation of two new CME models in EUHFORIA. Both models possess a toroidal geometry, but the internal distribution of the magnetic field is different.}
 {We introduce the two toroidal CME models analytically, along with their numerical implementation in EUHFORIA. One model is based on the modified Miller-Turner (mMT) solution, while the other is derived from the Soloviev equilibrium, a specific solution of the Grad-Shafranov equation. The magnetic field distribution in both models is provided in analytic formulae, enabling a swift numerical computation. After detailing the differences between the two models, we present a collection of thermodynamic and magnetic profiles obtained at Earth using these CME solutions in EUHFORIA with a realistic solar wind background. Subsequently, we explore the influence of their initial parameters on the time profiles at L1. In particular, we examine the impact of the initial density, magnetic field strength, velocity, and minor radius.}
 {The Soloviev model allows control over the shape of the poloidal cross section, as well as the initial twist. In EUHFORIA, we obtained different thermodynamic and magnetic profiles depending on the CME model used. The generated magnetic profiles reflect the initial magnetic field distribution of the chosen model. We found that changing the initial parameters affects both the amplitude and the trend of the time profiles. For example, using a high initial speed results in a fast evolving and compressed magnetic structure. The speed of the CME is also linked to the strength of the initial magnetic field due to the contribution of the Lorentz force on the CME expansion. However, increasing the initial magnetic field also increases the computation time. Finally, the expansion and integrity of the magnetic structure can be controlled via the initial density of the CME.}
 {Both toroidal CME models are successfully implemented in EUHFORIA and can be utilized to predict the geo-effectiveness of the impact of real CME events. Moreover, the current implementation could be easily modified to model other toroidal magnetic configurations.} 



 \maketitle
%
\section{Introduction} \label{sec:Introduction}

In the solar corona, rapid changes in the magnetic configuration can lead to the release of substantial quantities of magnetic energy, a phenomenon known as a solar flare \citep{Forbes06,Schmieder2007}. In some cases, flares can also be accompanied by plasma ejections, e.g.\ filaments, jets \citep{Schmieder1997,Schmieder2013}. We refer to these as eruptive flares (as opposed to confined flares) \citep{Zuccarello2017}. If the ejected structure extends spatially and is subsequently observed in the high corona in white light using a coronagraph, it defines a Coronal Mass Ejection (CME). These spectacular phenomena can impact Earth's magnetic environment and human technologies \citep{Bothmer07}. Among other effects, satellite communications can be disrupted, GPS signals can be lost due to electronic variations in the ionosphere, and geomagnetically induced currents can damage the power grid and affect the erosion of oil pipelines. 

The geoeffectiveness of a solar transient depends on its hydrodynamic and magnetic properties \citep{Dumbovic15}. Signatures of Interplanetary Coronal Mass Ejections (ICMEs) can be measured by different spacecraft such as ACE \citep{Smith98}, the Parker Solar Probe \citep{Fox16}, STEREO A and B \citep{Kaiser07}, the Solar Orbiter \citep{Muller20}, Helios \citep{Roberts87}, and Wind \citep{Bougeret95}. For some events, in situ measurements reveal an interplanetary shock of the ICMEs, followed by a heated region called the sheath as a result of the accumulation of matter upstream of the expansion of the magnetic ejecta \citep[ME; ][]{Kaymaz06}. The ME, characterized by a strong magnetic signature and low temperature, is the direct signature of the twisted structures ejected from the solar corona. This structure is usually described as a flux rope \citep{Demoulin08}. While interplanetary shocks alone can trigger magnetic disturbances on Earth, the strongest geomagnetic storms occur when the $B_z$ component of the magnetic field of the ME, i.e.\ the component of the magnetic field perpendicular to the equatorial surface,  has a direction opposite to that of the magnetopause \citep{Lugaz16}. Such a configuration allows for reconnection between the magnetic field lines of the two magnetic structures, which leads to an energy transfer towards the inner regions of the magnetosphere \citep{Akasofu81,Dungey61}.

ICMEs present complex signatures that reflect their interaction with the solar wind during their propagation in the heliosphere \citep{Scolini22}. For example, using a 2.5D MHD simulation, \citep{Zuccarello11} found that the CME can be deflected toward the current sheet of a large streamer due to a discrepancy in the magnetic pressure and tension forces. Recently, \citet{Asvestari22} found that the CME tilt evolves, depending on the strength and orientation of the ambient magnetic field. Interactions with the surrounding field can also lead to rotation of the CME \citep[e.g.,][]{Liu18,Manchester17,Shiota10}. It should also be noted that the thermodynamic and magnetic properties of successive CMEs are directly affected by their interactions with each other \citep{Scolini20,Koehn22}.

Although multiple viewpoint reconstruction techniques are beginning to reconstruct the complex geometric and magnetic structure of a CME from the current set of observations \citep[e.g.,][]{Rodari18}, the current limited number of satellites does not enable us to fully determine all the properties of ICMEs during their propagation \citep{Demoulin10}. Therefore, in addition to observations, a set of numerical codes has been developed with the aim of tracking the evolution of CMEs in the heliosphere. In a space weather context, their use is particularly relevant for attempting to predict the geoeffectiveness of a CME before it reaches Earth. Examples of such 3D magnetohydrodynamics (MHD) simulations include ENLIL \citep{Odstrcil03}, EUHFORIA \citep{Pomoell2018}, MS-FLUKSS \citep{Singh18}, SUSANOO-CME \citep{Shiota16}, and ICARUS \citep{Verbeke22,Baratashvili22}, where the latter uses the versatile MPI-AMRVAC framework \citep{Keppens2023} as its MHD solver, enabling radial grid stretching and solution adaptive mesh refinement. In these various simulation models, one or more CMEs propagate through a realistic background solar wind after their injection at the inner boundary of the domain (usually located at 0.1~AU). The predictions are obtained by extracting the temporal evolution of the plasma quantities (velocity, pressure, density, magnetic field, etc.) at different points in the heliosphere. The obtained time profiles depend, especially at L1/Earth, on the inserted CME model and its magnetic and thermodynamic properties (density, speed, magnetic flux, etc.). The characteristics of the injected CME can be deduced from remote sensing observations \citep[e.g.,][]{Scolini19,Maharana22}.

In this work, we are particularly interested in the 3D time-dependent magnetohydrodynamic (MHD) model of heliospheric wind and CME evolution, EUHFORIA \citep[European Heliospheric Forecasting Information Asset][]{Pomoell2018}. Before this study, three CME models, each with its own capabilities and limitations, were implemented in the framework of EUHFORIA. The simplest is the cone model, which represents an unmagnetized plasma with a self-similar expansion \citep{Xie04}. In EUHFORIA in combination with an energetic particle acceleration and transport model called PARADISE, the cone model can be used to model the interplanetary acceleration of low-energy protons during an energetic storm particle (ESP) event \citep[][]{Wijsen22}. However, this CME model is obviously not suitable for studying the evolution of the internal magnetic signature of an ICME. The second model implemented in EUHFORIA is the spheromak model, representing a CME with a global spherical shape but with an initially linear force-free magnetic field \citep{Chandrasekhar57,Verbeke19}. Although the use of this magnetized CME model is widespread \citep[e.g.,][]{Verbeke19,Scolini19,Asvestari22,Verbeke22}, as a result of its spherical geometry lacking CME legs, the spheromak model does not adequately model scenarios where Earth is impacted by the flanks/legs of ICMEs. To address this limitation, \citet{Maharana22} implemented in EUHFORIA a flux rope model in which the CME has an extended flux-rope geometry including CME legs \citep[e.g.,\ "Flux Rope in 3D", or FRi3D;][]{Isavnin16}. Its geometry is much closer to observations where CMEs are often observed to possess crescent-shaped shapes \citep{Janvier13}. However, solving the complex equations defining the magnetic distribution of the model requires a substantial amount of computing time, which limits the use of this model in an operational space weather context.

In this study, we introduce the implementation of two new CME models with toroidal geometry in EUHFORIA, with the aim of overcoming the shortcomings of the current CME models. The toroidal geometry of the new CME models is simpler than that of the realistic FRi3D model. However, unlike the latter, we suggest that the torus models are better suited for operational space weather forecasting. Indeed, their magnetic and geometric configurations are based on straightforward analytical formulations. Consequently, they can be applied as time-dependent boundary conditions at 0.1 AU, making them much more time-efficient than the numerically determined FRi3D model.

On the other hand, although the new models do not remain connected to the Sun after injection, just like the spheromak model, they possess a geometry that is more consistent with the observations than the spheromak. Indeed, the toroidal models have a curved front and two 'legs', in contrast to the spheromak. Additionally, one of our models is based on a particular solution of the Grad-Shafranov equation. This equation is frequently employed to determine the cross-section of observed magnetic clouds \citep{Mostl09,Isavnin11}. Hence, we anticipate simulations with the new toroidal CME models would closely resemble the cross-sections of the observed ones. Furthermore, in this particular model, it is possible to adjust the ratio of the poloidal and toroidal fluxes. This feature is absent in other models and provides an added degree of freedom to more accurately predict the magnetic profiles observed at Earth.

Lastly, while both the CME models introduced here exhibit a toroidal geometry, they differ in the distribution of the internal magnetic field. This means we can expect varying magnetic field profiles in EUHFORIA, underscoring the rationale for implementing both models. It is important to note that this work focuses on validating the numerical implementation of these two models. In particular, we examine how the models' distinct free parameters impact the time profiles recorded at Earth (extracted from the 3D time-dependent simulations). While addressing all potential configurations is not feasible, our research seeks to broadly demonstrate the potential of these two new models in EUHFORIA. A subsequent study will be devoted to directly comparing them with the FRi3D and spheromak models to highlight the strengths and weaknesses of all numerical CME models.

The paper is organized as follows: Section~\ref{sec:CMEmodel} introduces the two new torus CME models : one with the modified Miller-Turner solution \citep{Vandas15}, and the other with the Soloviev solution \citep[an analytical solution of the Grad-Shafranov equation; ][]{Soloviev75}. In Section~\ref{sec:implementationboth}, we provide details on the implementation of the new models in the framework of EUHFORIA. The resulting thermodynamic and magnetic field profiles associated with the CME models in the heliosphere are discussed in Section~\ref{sec:thermo_mag_profile}. In this section, the influence of various free parameters of the CME models is also investigated. We specifically discuss the impact of initial radial speed (cf.\ Sect.~\ref{sec:speed}), magnetic field strength (cf.\ Sect.~\ref{sec:B0}), density (cf.\ Sect.~\ref{sec:density}), as well as minor radius (cf.\ Sect.~\ref{sec:impacta}). To conclude, Section \ref{sec:conclusion} summarizes the main findings and potential implications for future research.

\section{CME models} \label{sec:CMEmodel}
We implement in EUHFORIA two different CME models. In this section, we provide an overview of the two CME models: (1) the modified Miller-Turner solution \citep{Vandas15}, and (2) the Soloviev solution, an analytical solution of the Grad-Shafranov equation \citep{Soloviev75}.

\subsection{The Modified Miller-Turner solution}\label{sec:Millerturner}
\begin{figure}[t!]
    \centering
    \includegraphics[width=0.5\textwidth]{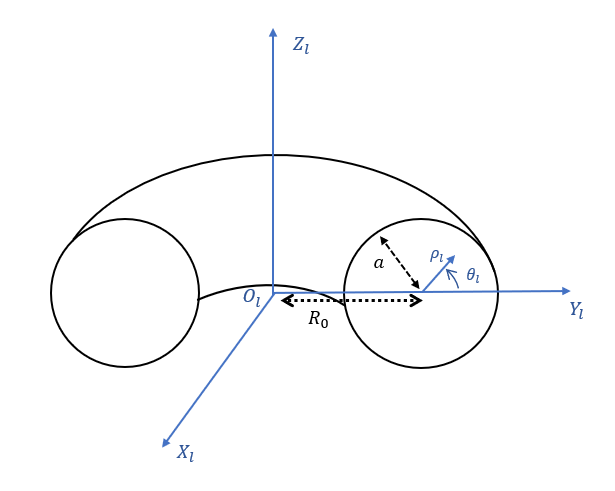}
    \caption{Local toroidal curved cylindrical coordinates associated with the Miller-Turner CME model. In the local Cartesian system centered on $O_l$, the $z$-axis coincides with the symmetry axis of the torus. }
    \label{fig:systemturner}
\end{figure}

The magnetic configuration of this model was originally derived by \cite{Miller81} and represents a constant-$\alpha$ force-free field in a toroidal geometry given by:
\begin{eqnarray} 
B_{\rho_{l}}&=& \frac{B_{0}}{2\alpha_{MT} R_{0}}J_{0}(\rho_{l}\alpha_{MT})\sin\theta_{l} \,, \label{eq:orMT_1} \\
B_{\phi_{l}}&=& B_{0}\left(1-\frac{\rho_{l}}{2 R_{0}}\cos\theta_{l}\right)J_{0}(\rho_{l}\alpha_{MT}) \,,\label{eq:orMT_2} \\
B_{\theta_{l}}&=& B_{0}\frac{\cos\theta_{l}}{2\alpha_{MT} R_{0}}J_{0}(\rho_{l}\alpha_{MT}) \nonumber \\
&&-B_{0}\left(1-\frac{\rho_{l}\cos\theta_{l}}{2R_{0}}\right)J_{1}(\rho_{l}\alpha_{MT}) \,,
\label{eq:orMT_3}
\end{eqnarray}
where $(\rho_{l},\phi_{l}, \theta_{l})$ are the local toroidal curved cylindrical coordinates, $a$ the minor radius of the torus, $R_{0}$ the major radius and $B_{0}$ the strength of the toroidal component in the center of the torus (cf.\ Fig.~\ref{fig:systemturner}). $J_{0}$ and $J_{1}$ are the Bessel functions of the first kind of order 0 and 1, respectively. The local angles $\phi_{l}$ and $\theta_{l}$ range from $0$ to $2\pi$, while $\rho_{l}$ covers the interval $[0,a]$. Hence, this CME model has a fixed circular poloidal cross-section unlike the other implemented model which contains parameters that determine the shape of the poloidal cross-section (cf.\ Sect.~\ref{sec:Soloviev_theory}). To ensure that the magnetic field is completely poloidal (i.e.,\ $B_{\phi_{l}}=0$) at the boundary of the flux rope and that the magnetic field line is confined in the torus, the value $\alpha_{MT}$ is such that $a\alpha_{MT} $ is the first root of $J_{0}(\rho_{l}\alpha_{MT})$ : 
\begin{equation}
    \alpha_{MT}\approx C_{\alpha}\frac{2.41}{a}, 
\end{equation}
where $C_{\alpha}$ is defined as the chirality of the magnetic field and takes value $+1$ ($-1$) for right (left) handedness. The Miller-Turner solution tends to the Lundquist solution \citep{Lundquist51} when $R_{0} \to \infty$. 

This solution has one main disadvantage: the solenoidal condition $\nabla\cdot \mathbf{B} = 0$ is not exactly fulfilled, but only approximately. To mitigate the effect of the non-solenoidal scenario, in this work, we use a modified Miller-Turner (mMT) solution defined by \cite{Romashets03} such that:
\begin{equation}
    \mathbf{B}=\nabla\times\frac{\mathbf{B}_{MT}}{\alpha_{MT}},
\end{equation}
where $\mathbf{B}_{MT}$ is the original Miller-Turner solution (cf.\ Eqs ~\ref{eq:orMT_1}-~\ref{eq:orMT_3}). The new magnetic field can be expressed as : 
\begin{eqnarray} 
B_{\rho_{l}}&=& B_{0}\frac{R_{0}-2\rho_{l}\cos\theta_{l}}{2\alpha_{MT} R_{0}(R_{0}+\rho_{l}\cos\theta_{l})}J_{0}(\rho_{l}\alpha_{MT})\sin\theta_{l} \,,\label{eq:newMT_1} \\
B_{\phi_{l}}&=& B_{0}\left(1-\frac{\rho_{l}}{2 R_{0}}\cos\theta_{l}\right)J_{0}(\rho_{l}\alpha_{MT}) \,,\label{eq:newMT_2} \\
B_{\theta_{l}}&=& B_{0}\frac{R_{0}-2\rho_{l}\cos\theta_{l}}{2\alpha_{MT} R_{0}(R_{0}+\rho_{l}\cos\theta_{l})}J_{0}(\rho_{l}\alpha_{MT})\cos\theta_{l} \nonumber\\ 
&&-B_{0}\left(1-\frac{\rho_{l}}{2 R_{0}}\cos\theta_{l}\right)J_{1}(\rho_{l}\alpha_{MT}) \,.\label{eq:newMT_3}
\end{eqnarray}
In the mMT configuration, the magnetic field is fully divergence-free and the degree of twisting varies from zero at the center to an infinite value at the outer boundary.

\begin{figure}[t!]
    \centering
    \includegraphics[width=0.5\textwidth]{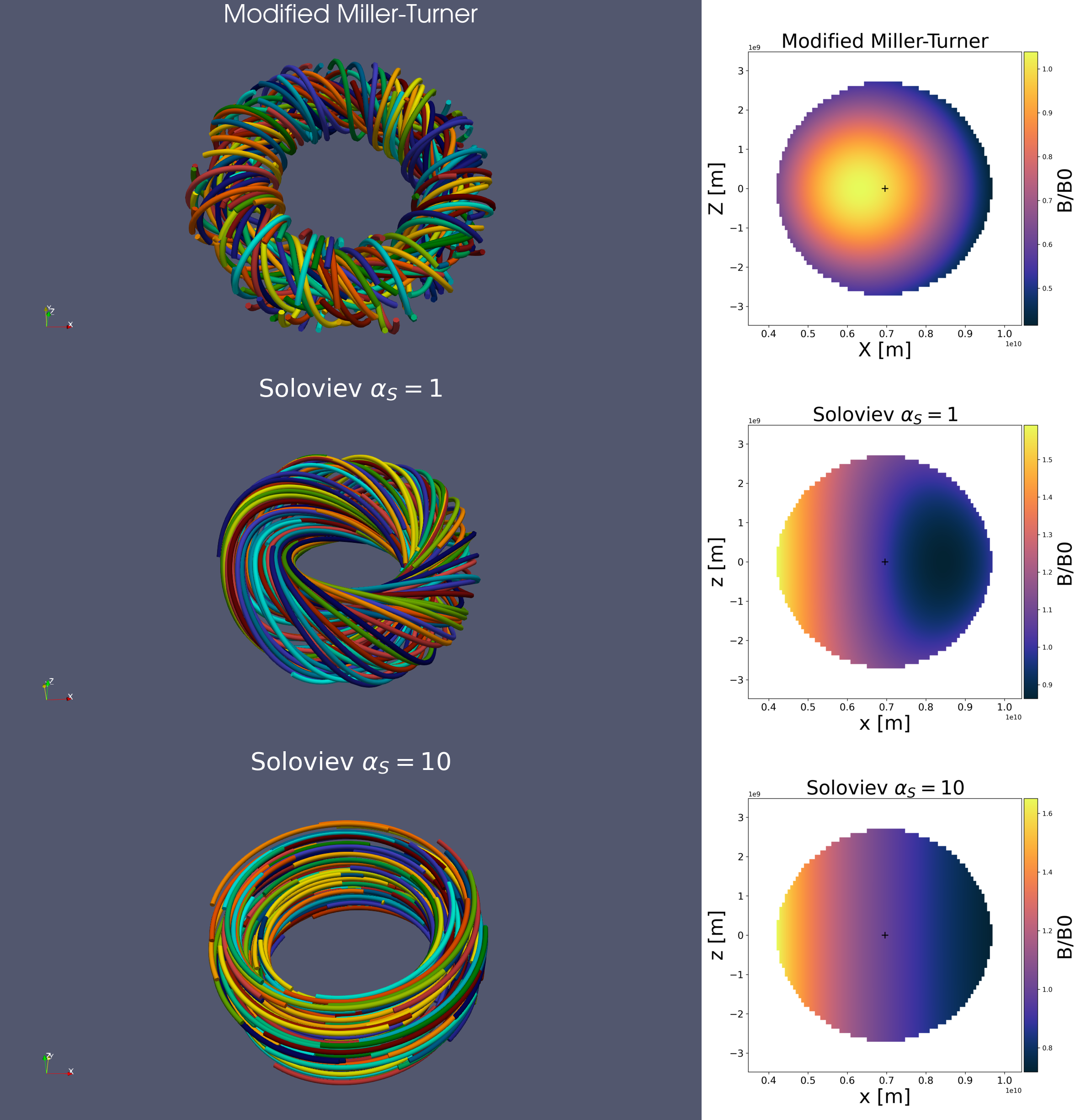}
    \caption{The normalized magnetic field strength in the two CME models implemented in EUHFORIA. The left panels show a sample of magnetic field lines in a torus with a minor radius $a=4\;R_{\odot}$ and a major radius $R_{0}=10\;R_{\odot}$. The right panels show the dimensionless magnetic field strength $B/B_{0}$ in a poloidal cross-section. The "+" markers indicate the center of the torus. From top to bottom: magnetic field configuration in the modified Miller-Turner model, in the Soloviev model with $\alpha_{S}=1$ and with $\alpha_{S}=10$. When the $\alpha_{S}$ parameter is too high, the magnetic field is mainly toroidal.}
    \label{fig:crossection}
\end{figure}

In Fig.~\ref{fig:crossection}, the top panels shows the magnetic field distribution in the mTM solution. In this figure, the center of the torus is at the origin $(0,0,0)$ of the local coordinate system, assuming that the rotation axis aligns with the $z$-axis. The major radius, $R_{0}$, is equal to $10\;R_{\odot}$, while the minor radius is $a=4\;R_{\odot}$. The magnetic field in the circular cross-section has a maximum shift toward the torus hole. According to \citet{Vandas15}, the shift increases when the aspect ratio, $R_{0}/a$, decreases. In theory, the magnetic field is fully force-free only for large aspect ratios ($R_{0}/a\gg1$). However, \citet{Vandas15} found that the force-free condition in the original Miller-Turner solution (cf.\ Eqs.~\ref{eq:orMT_1}-\ref{eq:orMT_3}) is held much better than in the mMT model for low aspect ratios ($\approx2$). In this study, we have chosen the mMT solution, which is unique in being completely divergence-free, thus avoiding numerical instability and unphysical solutions. It is worth noting that there is also the Tsuji solution, which describes a magnetic configuration in the same toroidal geometry but is force-free for all aspect ratios \citep{Tsuji91}. That being said, the expressions for its magnetic field components contain infinite series expansions and the coordinate system transformations required for the implementation in EUHFORIA are too complex for a straightforward and quick numerical implementation.

\subsection{The Soloviev solution}\label{sec:Soloviev_theory}

\subsubsection{Analytical formulation}

\begin{figure}[t!]
    \centering
    \includegraphics[width=0.5\textwidth]{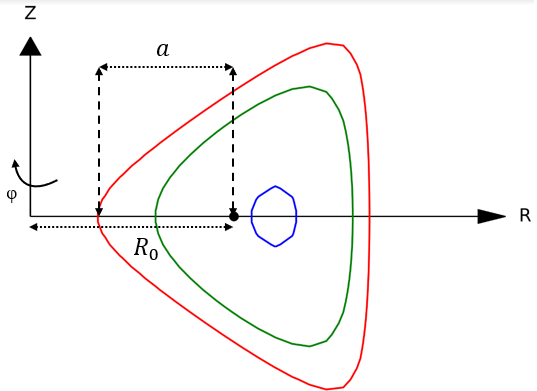}
    \caption{Particular cylindrical system associated with the Soloviev solution. The color lines are flux isocontours such as $\psi=1$ (in red), $\psi=0.7$ (in green) and $\psi=0.2$ (in blue).}
    \label{fig:systemsoloviev}
\end{figure}

The second CME model originally corresponds to a global magnetic confinement topology in a tokamak configuration \citep{Goedbloed84}. This configuration can be described by a particular cylindrical system with $R$ representing the distance from the center to the symmetry axis, $Z$ the vertical coordinate, and $\phi$ the toroidal angle (cf.\ Fig.~\ref{fig:systemsoloviev}). Considering we assume axisymmetry along $\phi$, and that the static equilibrium solution satisfies the following MHD equations
\begin{equation}
\textbf{J}\times\textbf{B}=\nabla p \text{,} \qquad
\textbf{J} =\nabla \times \textbf{B}, \qquad \text{and} \qquad
\nabla\cdot\textbf{B}=0 \,, 
\end{equation}
we find that all solutions are described by the non-linear Grad-Shafranov equation
\begin{equation}
\Delta\Psi - \frac{1}{R}\frac{\partial\Psi}{\partial R} = R\frac{\partial}{\partial R}\left(\frac{1}{R}\frac{\partial\Psi}{\partial R}\right)+\frac{\partial^{2}\Psi}{\partial Z^{2}}=-II'-R^{2}p'\text{.} \label{eq:Grad}
\end{equation}
Here, $\Psi$ is the poloidal flux, which ranges from $0$ on the magnetic axis to $\Psi_{1}$ on the plasma boundary. The Grad-Shafranov equation, Eq.~\ref{eq:Grad}, depends on two arbitrary flux functions: the stream function $I(\Psi)$ of the poloidal current and the pressure $p(\Psi)$ \citep[cf.\ Section 16.2][for the derivation]{Goedbloed19}.

The magnetic configuration used in our work is obtained from an exact analytical solution of the Grad-Shafranov equation, called the Soloviev equilibrium. This particular solution imposes linear profiles for $I^{2}$ and $p(\Psi)$ (i.e., for the case when the right-hand side of Eq.~\ref{eq:Grad} is a constant). The dimensionless flux $\psi=\Psi/\Psi_{1}$ is then given by: 

\begin{equation}
    \psi = \Bigg[X - \frac{1}{2}\epsilon(1-X^2)\Bigg]^2 +\Bigg(1-\frac{\epsilon^{2}}{4}\Bigg) \Bigg[1+\epsilon\tau X(2+\epsilon X)\Bigg]\Bigg(\frac{Y}{\sigma}\Bigg)^2 ,
\label{eq:phi_soloviev}
\end{equation}
which depends on the dimensionless poloidal coordinates
\begin{eqnarray}
    X&=&\frac{R-R_{0}}{a},\;\; \text{ \ and}\\
    Y&=&\frac{Z}{a}\,,
\end{eqnarray}
as well as the inverse aspect ratio $\epsilon=a/R_{0}$ ($\ll 1$ in asymptotic expansions), the triangularity $\tau$, and the elongation (or ellipticity)  $\sigma$. Using the dimensionless quantity $\psi$, the boundary condition becomes $\psi(X,Y)=1$ on the outer boundary of the plasma cross-section.

\begin{figure}[h!]
    \centering
    \includegraphics[width=0.4\textwidth]{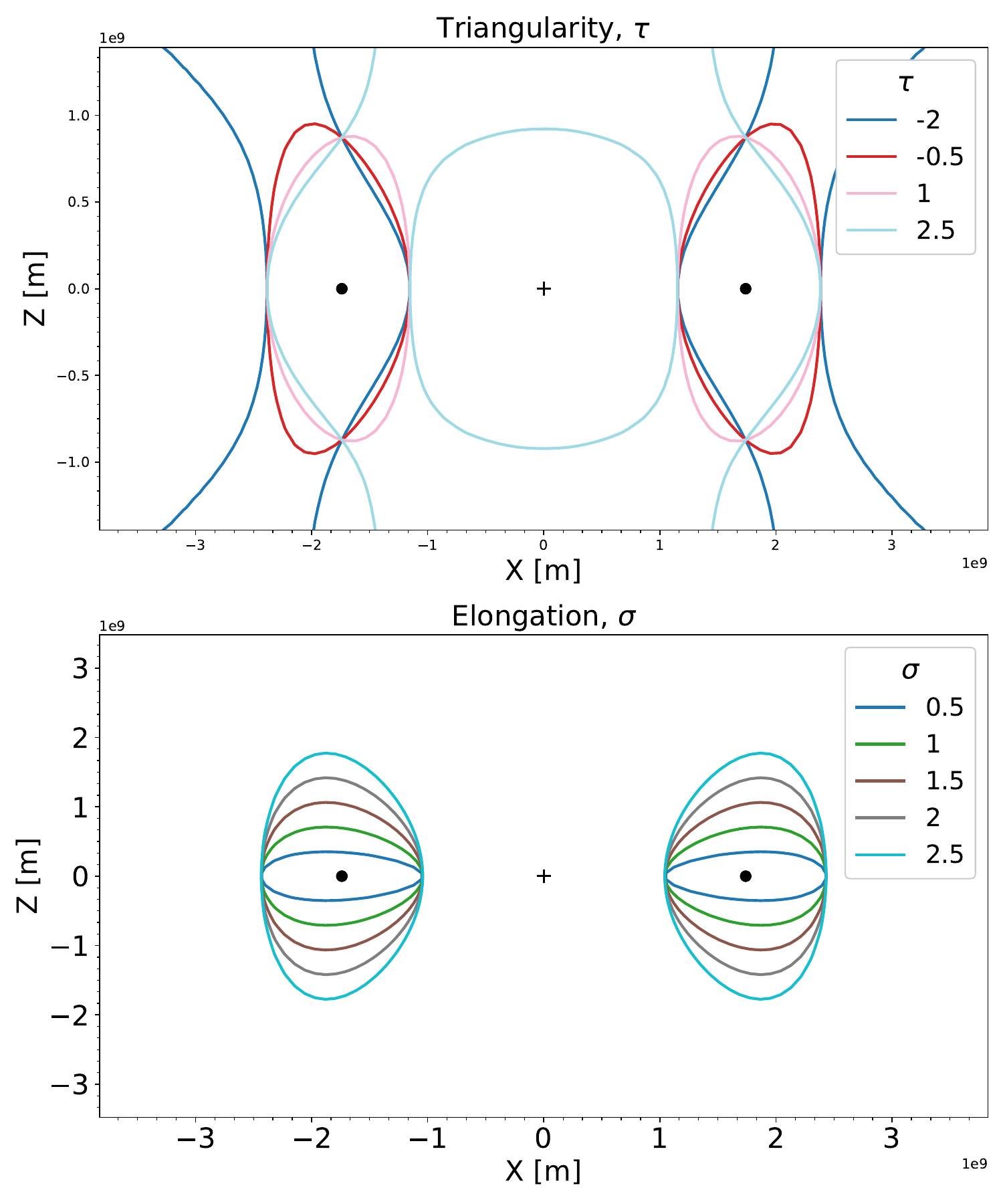}
    \caption{Isocontours of $\psi=1$ in the $X-Z$ plane for the Soloviev solution with different geometries. The upper panel illustrates the impact of the triangularity parameter $\tau$ on the isocontour $\psi=1$, while the lower panel, characterizes the influence of the elongation parameter $\sigma$ for a torus with $a=1~\;R_{\odot}$ and $R_{0}=2.5~\;R_{\odot}$, centered at (0,0,0). The marker "+" indicates the center of the torus, while the dot marker corresponds to $x=R_{0}$. In the top panel, $\sigma$ is kept constant at $1.4$ and in the bottom plot, $\tau$ is constrained to $0$.}
    \label{fig:contour}
\end{figure}

\subsubsection{Geometry} \label{sec:geometrysoloviev}
In the Miller-Turner model, the poloidal cross section is circular (cf.\ the top panels, Fig.~\ref{fig:crossection}). However, it is possible to change the shape of the poloidal cross-section in the Soloviev solution by varying the triangularity $\tau$ and the elongation $\sigma$ parameters. Figure~\ref{fig:contour} shows the isocontour $\psi=1$ in the Soloviev model as a function of triangularity and elongation. 

The shape of the flux contour can also be modified by changing the triangularity $\tau$ (cf.\ Fig.~\ref{fig:contour}, top panel). The triangularity parameter allows one to pass from an elliptic flux contour when $\tau=1$ (and $\sigma>1$) to a triangular flux contour whose direction depends on the sign of the triangularity (cf.\ Fig.~\ref{fig:contour}, top panel). For $\tau=-0.5$ and $\tau=1$ (cf.\ Fig.~\ref{fig:contour}, top panel), the isocontour $\psi=1$ is closed and delineates the limits of the torus. However, the flux contour is open for $\tau=-2$ and $\tau=2.5$, which means that the torus is not properly delineated. Analytically, this flux contour shape reflects the presence of hyperbolic points located at : 
\begin{eqnarray}
    x_{s}&=&-\frac{1}{\epsilon}\left[1-\frac{1}{\tau}\right],\\
    y_{s}&=&\pm\frac{\sigma}{\epsilon\tau}\sqrt{\frac{1}{2}\frac{1+\epsilon^{2}\tau}{1-\frac{\epsilon^{2}}{4}}}, 
\end{eqnarray}
that are present in the domain if $\tau<-1/[\epsilon(2+\epsilon)]$ or $\tau>1/[\epsilon(2-\epsilon)]$. To obtain a closed flux contour, the value of $\tau$ must be restricted within these limits. For example, with an inverse aspect ratio of $0.4$ as in Fig.~\ref{fig:contour}, the triangularity must lie within the range $[-1.04, 1.56]$ to obtain a closed contour. The values $-2$ and $2.5$, being outside this range, result in an open isocontour for $\psi=1$ in both cases, as shown in Fig.~\ref{fig:contour}. 

The elongation $\sigma$ does not change the size in the toroidal plane of the torus, but affects the cross-sectional shape of the torus along the $Z$-direction, as shown in Fig.~\ref{fig:contour}. Indeed, the outer isocontour moves from a near-circular cross section when $\sigma=1$ to an ellipsoidal cross section when $\sigma>1$ (cf.\ Fig.~\ref{fig:contour}, bottom panel). The maximum height reached in the $Z$-direction is $\sigma a$. For further information on the geometry of the torus, we refer to \citet{Goedbloed84}.

\subsubsection{Magnetic field configuration}

Using the analytical solution for the flux (cf.\ Eq.~\ref{eq:phi_soloviev}), the magnetic field components in the particular cylindrical system can be determined such as : 
\begin{eqnarray}
B_{R}&=&-\frac{B_{0}}{\alpha_{S}}\frac{\epsilon}{1+\epsilon X}\frac{\partial\psi}{\partial Y} \,,\label{eq:BR}\\
B_{Z}&=&\frac{B_{0}}{\alpha_{S}}\frac{\epsilon}{1+\epsilon X}\frac{\partial\psi}{\partial X} \,,\label{eq:BZ}\\
B_{\phi}&=& \frac{B_{0}}{1+\epsilon X}\left[1-2\frac{\epsilon\psi}{\alpha_{S}^{2}}\left(A\epsilon-\frac{B}{2}\right)\right]^{1/2} \,,\label{eq:Bphi}
\end{eqnarray}
where,
\begin{eqnarray}
A&=&2\left[1+\frac{1-\frac{\epsilon^{2}}{4}}{\sigma^{2}}\right]\text{,} \\
B&=&4\epsilon\left[1+\frac{1-\frac{\epsilon^{2}}{4}}{\sigma^{2}}\tau\right]\text{,} 
\end{eqnarray}
and
\begin{equation}
    \frac{\partial\psi}{\partial X}=[X-(1-X^2)](1+\epsilon X)+\bigg(1-\frac{\epsilon^2}{4}\bigg)[2\epsilon\tau(1+\epsilon X)]\left(\frac{Y^2}{\sigma^2}\right) \text{,}
\end{equation}
while 
\begin{equation}
    \frac{\partial \psi}{\partial Y}=\bigg[1-\frac{\epsilon^2}{4}\bigg]
[1+\epsilon\tau X(2+\epsilon X)]\left(\frac{2Y}{\sigma^2}\right)\,,
\end{equation}
and $\alpha_{S}$ is related to the inverse of the total poloidal flux through the plasma and defined as 
\begin{equation}
    \alpha_{S}=\frac{a^{2}B_{0}}{\Psi_{1}} \text{.} \label{eq:alpha}
\end{equation}
Here, $B_{0}$ is a constant magnetic field strength used to scale the different magnetic field components (cf.\ Eqs.~\ref{eq:BR}, ~\ref{eq:BZ}, and ~\ref{eq:Bphi}).

Contrary to the modified Miller-Turner CME model (cf.\ Sect.~\ref{sec:Millerturner}), the Soloviev model allows for varied magnetic field distributions for the same poloidal flux $\Psi_{1}$, depending on the value of the parameter $\alpha_{S}$. In the Miller-Turner CME model, the magnetic helicity, which signifies the winding of magnetic field lines relative to one another, is a constant parameter. However, the Soloviev solution introduces flexibility in setting helicity by adjusting the parameter $\alpha_{S}$ (cf.\ Eq.~\ref{eq:alpha}). Indeed, the local magnetic fields $B_{R}$ and $B_{Z}$ (cf.\ Eqs.~\ref{eq:BR} and \ref{eq:BZ}, respectively) exhibit variations proportional to $1/\alpha_{S}$, while the poloidal field $B_{\phi}$ presents a more intricate dependence (cf.\ Eq.~\ref{eq:Bphi}). As a result, the adjustment of $\alpha_{S}$ modifies the ratio between the toroidal and poloidal fields.

To visualize this, Fig.~\ref{fig:crossection} displays the magnetic field distribution in both the mMT solution and the Soloviev model with $\alpha_{S}=1$ and $\alpha_{S}=10$. A distinct variance can be observed in the distribution of magnetic field lines within the Soloviev model between the $\alpha_{S}=1$ and $\alpha_{S}=10$ cases. In the latter scenario, the high $\alpha_{S}$ value leads to the dominance of the magnetic field $B_{\phi}$ resulting in an almost purely toroidal magnetic field. It is worth noting that, in a tokamak configuration, when the toroidal magnetic field varies as $B_\phi=I(\Psi)/R$, this component is always dominant \citep{Goedbloed84}.

Moreover, the different magnetic helicities within the Soloviev and mMT models indicate their distinct magnetic field distributions. As seen in the right panels in Fig.~\ref{fig:crossection}, the isocontours of constant $B$ in the mMT model are circular, while in the Soloviev model they appear as near-vertical bands. It is also worth noting that the maximum of magnetic field is not located at the center, but rather at the boundary in the Soloviev model, which is not usually the case in the magnetic configurations popularly used to model CMEs \citep{Lundquist51,Chandrasekhar57}.  Given these disparities in magnetic field distributions between the Soloviev and the mMT models, we aim to validate the implementation of both models in EUHFORIA and investigate the differences in their profiles in the heliosphere.

\section{Numerical implementation of Miller Turner CME in EUHFORIA} \label{sec:implementationboth}

\subsection{EUHFORIA}

The two magnetic configurations used are implemented as CME models in the physics-based 3D magnetohydrodynamic simulation of the heliosphere wind and CME evolution, EUHFORIA \citep[European Heliospheric Forecasting Information Asset][]{Pomoell2018}. For space weather operational purposes, EUHFORIA can be used to track the propagation of one or more CMEs in the heliosphere and predict the temporal evolution of different plasma quantities (e.g.\ density, velocity, magnetic field) at different positions in the inner heliosphere.

EUHFORIA is divided into two main parts. The first one is the semi-empirical Wang–Sheeley–Arge model \citep[WSA][]{McGregor11,van10} that inputs a synoptic magnetogram to provide the plasma quantities at 0.1~AU needed to establish the background solar wind in the heliosphere. The magnetogram can come from different sources, for example from the Global Oscillation Network Group \citep[GONG; ][]{Harvey96} or from the SDO's Helioseismic and Magnetic Imager \citep[HMI; ][]{Schou12}. In this coronal part, the 3D coronal magnetic field is first computed from a Potential Field Source Surface (PFSS) extrapolation \citep{Altschuler69}. The magnetic field is then extended to 0.1~AU using the Schatten Current Sheet model \citep{Schatten69}. The solar wind speed at the 0.1~AU boundary is obtained as a function of the flux tube expansion factor, $f$,  and the distance, $d$, of the foot point of the flux tube to the nearest coronal hole. From this solar wind speed $v_{sw}=v(f,d)$, the density and the temperature at the outer boundary are computed using empirical relations \citep[cf.][for more details]{Pomoell2018}.

The second part of EUHFORIA is a 3D time-dependent magnetohydrodynamic simulation of the heliosphere. The thermodynamic and magnetic quantities, computed from the coronal model, are self-consistently determined between 0.1~AU and 2~AU in an MHD relaxation phase in a uniform grid. This evolution is determined by solving the set of ideal MHD equations with a polytropic index equal to 1.5, using a finite volume numerical architecture and a constrained transport scheme ensuring that the solenoidal condition is always satisfied up to machine accuracy.

EUHFORIA can be used to track the propagation of a CME and its possible interaction with the background solar wind by injecting a CME model at the inner boundary as described in the following section (cf.\ Sect.~\ref{sec:implementation}).

\subsection{CME implementation} \label{sec:implementation}
\begin{figure}[t!]
    \centering
    \includegraphics[width=0.4\textwidth]{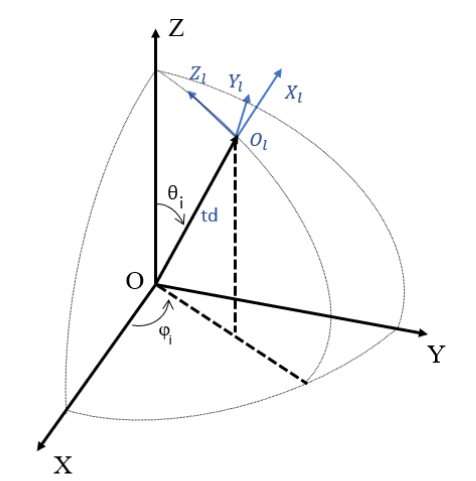}
    \caption{Relation between the global Cartesian coordinate system of EUHFORIA ($X,Y,Z$), and the local Cartesian system in which the two CME models are defined ($X_{l},Y_{l},Z_{l}$). The distance from the center $t_{d}$, the colatitude $\theta_{i}$, and the longitude $\phi_{i}$ define the initial position of the center of the torus, $O_{l}$. The vector $OO_{l}$, where $O$ is the center of the domain, corresponds to the $X_{l}$ axis of the local Cartesian coordinate system.}
    \label{fig:system}
\end{figure}

\begin{table*}[h!]
\centering
\begin{tabular}{|c|c|c|c|}
\hline
{Variable} &  {Description} & {Model} & {Discussion} \\ \hline
  {$t_{0}$} &  {Insertion time : date from which the torus is inserted into the domain} &  {Soloviev \& mMT}  & {-} \\ \hline
  {$\theta_{i}$} &  {Initial colatitude of the center} &  {Soloviev \& mMT} & {-}  \\ \hline
  {$\phi_{i}$} &  {Initial longitude of the center} &  {Soloviev \& mMT}  & {-}  \\ \hline
  {$a$} &  {Minor radius of the torus} &  {Soloviev \& mMT}  & {Sect.~\ref{sec:impacta}} \\ \hline
  {$R_{0}$} &  {Major radius of the torus} &   {Soloviev \& mMT} & {Sect.~\ref{sec:impacta}}  \\ \hline
  {$\tau$} &  {Triangularity} &  {Soloviev}  & {-} \\ \hline
  {$\sigma$} &  {Elongation} &  {Soloviev}  & {-} \\ \hline
  {$\omega$} &  {Angle defining rotation along the local axis $X_{l}$ } &  {Soloviev \& mMT}  & {-} \\ \hline
  {$v_{r0}$} &  {Initial radial speed of the center} &  {Soloviev \& mMT}  & {Sect.~\ref{sec:speed}}  \\ \hline
  {$n$} &  {CME density} &  {Soloviev \& mMT}  & {Sect.~\ref{sec:density}} \\ \hline
 {$T$} &  {CME Temperature} &  {Soloviev \& mMT} & {-}  \\ \hline
  {$B_{0}$} &  {Magnetic field strength} &  {Soloviev \& mMT}  & {Sect.~\ref{sec:B0}}  \\\hline
  {$\alpha_{s}$} &  {Twist} &  {Soloviev} & {Sect.~\ref{sec:reference}}  \\\hline
  {$C_{\alpha}$} &  {Chirality} &  {mMT}  & {Sect.~\ref{sec:reference}} \\
  \hline
\end{tabular}
\caption{Summary of the different free parameters defining the geometry, the magnetic and the kinetic properties of the implemented CMEs.  }
\label{tab:minimum}
\end{table*}

The CME injection in EUHFORIA takes place at the inner boundary of the heliospheric model, i.e.\ at 0.1~AU. Starting from a predetermined initiation time, at each time step increment a mask is computed to identify the points on the grid where the injected CME intersects the spherical boundary surface at 0.1~AU. At all these points, the solar wind speed, magnetic field, density, and temperature are substituted with those of the CME. In the subsequent time step, the CME center is advanced with the initially defined purely radial speed, $v_{r0}$, and the 3D mask region is updated as the solution moves through the spherical boundary.

Initially, the center $O_{l}$ of the torus is located at the position $(t_{d},\theta_{i},\phi_{i})$ in spherical coordinates, where $t_{d}$ is the distance from the center of the domain $O$ (0,0,0), $\theta_{i}$ is the initial colatitude and $\phi_{i}$ is the longitude (cf.\ Fig.~\ref{fig:system}). These three coordinates are free parameters of the models and must be set prior to the simulation.

At the intersection point, the components of the solar wind magnetic field that are replaced are in global spherical coordinates (${R}$, ${\theta}$, ${\phi}$). The equivalent global Cartesian system is ($X$, $Y$, $Z$) centered at $O$. Hence, it is necessary to transform the magnetic field computed according to Eqs.~\ref{eq:newMT_1}-\ref{eq:newMT_3} or \ref{eq:BR}-\ref{eq:Bphi} in a local toroidal coordinate system centered at $O_{l}$ into magnetic fields ($B_{R}$, $B_{\theta}$, $B_{\phi}$) in the global spherical system centered at $O$. The specifics of the base change are explained in Appendix~\ref{sec:appendix}. 

Different criteria are used to determine the intersection points for the Soloviev and the mMT models. In the case of the Soloviev solution, the flux $\psi$, is computed (cf.\ Eq.~\ref{eq:phi_soloviev}). All points on the boundary where the flux is less than or equal to 1 are considered to be within the CME. As mentioned in Sect.~\ref{sec:Soloviev_theory}, depending on the elongation, triangularity, and aspect ratio, the shape of the isocontour $\psi=1$ changes. In some cases, the isocontour is not closed (cf.\ Fig.~\ref{fig:contour}). Numerically, this will lead to the computation of a magnetic field that is not physical everywhere at the border. Therefore before using the Soloviev model, it must be analytically checked whether condition $\psi<1$ accurately defines a closed contour with the desired geometric parameters (cf.\ Sect.~\ref{sec:geometrysoloviev}).

For the Miller-Turner solution, the local toroidal radius $\rho_{l}$ (cf.\ Eq.~\ref{eq:rhol}) is computed; all points on the boundary where this radius is less than the minor radius of the torus are within the CME. This approach cannot be used for the Soloviev solution, as in that model the poloidal cross section is not (always) circular (cf.\ Sect.~\ref{sec:Soloviev_theory}).

In EUHFORIA, the two toroidal CME models are distinguished by their magnetic field configurations and the free geometrical parameters that define CME kinematics and kinetics. The CME parameters for both models are summarized in Table~\ref{tab:minimum}. In the following subsections, we will cover various parameters to highlight the differences between the two implemented CME models and how they affect the EUHFORIA simulation obtained at Earth. To limit the number of simulations, we will not discuss the impact of the initial position (i.e.,\ the angles $\theta_{i}$ and $\phi_{i}$) and the internal temperature of the CME ($T$). We also skip the triangularity and elongation parameters of the Soloviev model as they only influence simulations with high resolution and with large minor radius that enable clearly differentiating between a circular and a "triangular" cross section.

\section{Thermodynamic and magnetic profiles}\label{sec:thermo_mag_profile}

\subsection{Theoretical profiles} \label{sec:profileth}
\begin{figure*}[ht!]
    \centering
    \includegraphics[width=1\textwidth]{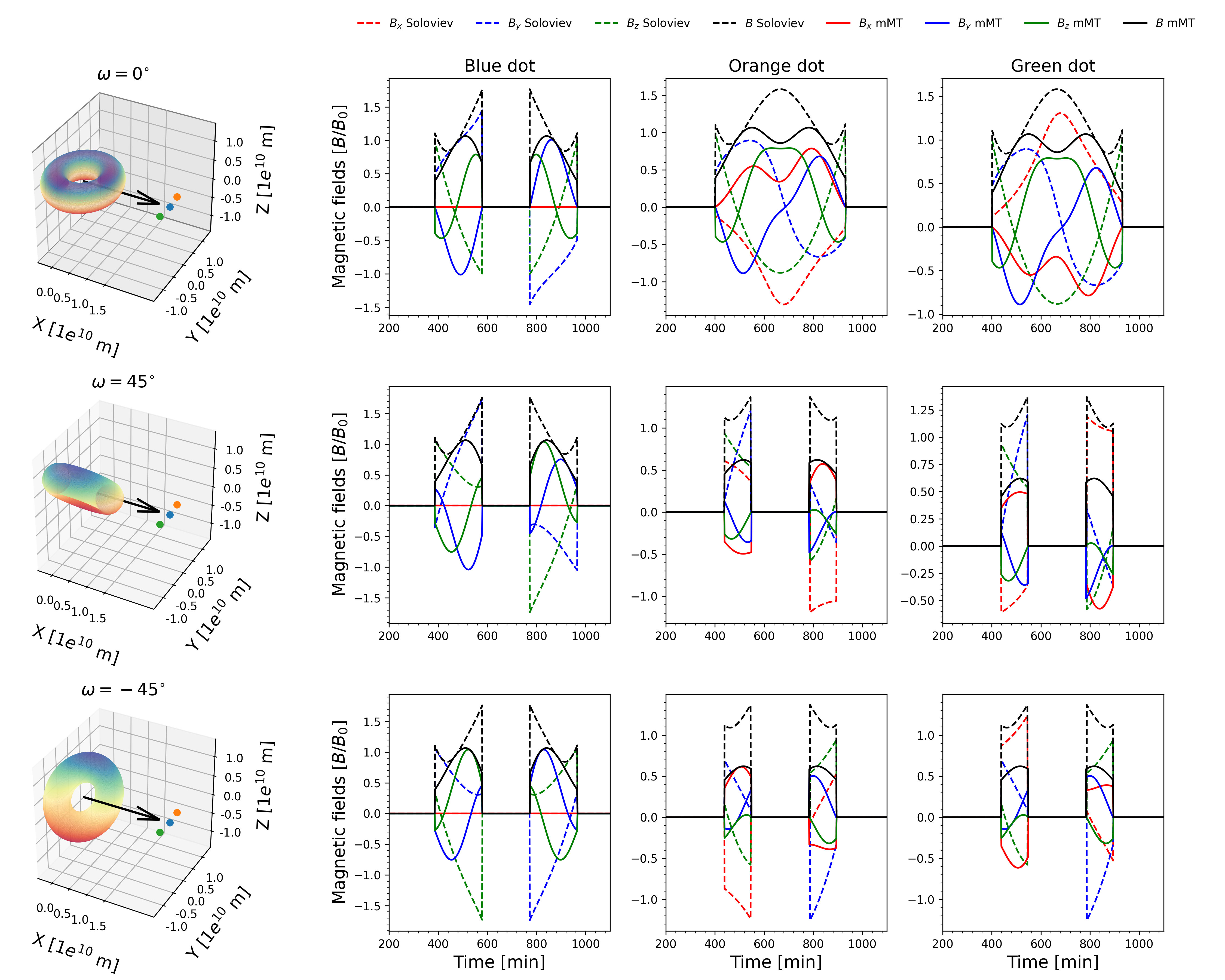}
    \caption{Theoretical profiles of magnetic field components are obtained by advancing a torus at a speed of $600$ km/s through three virtual satellites. The first column presents an isosurface $\psi=1$ for the Soloviev solution, with three torus inclinations: $\omega=0^{\circ}$ (top panel), $\omega=45^{\circ}$ (middle panel), and $\omega=-45^{\circ}$ (bottom panel). The second column represents, for these three configurations, the evolution of magnetic field components (in GSE) at the level of a virtual satellite (blue) placed directly in the direction of torus propagation. The third column corresponds to the passage through the offset virtual satellite (orange) at an angle $\phi=10^{\circ}$ relative to the propagation direction. The fourth column corresponds to the offset green satellite at an angle $\phi=-10^{\circ}$. The dashed lines correspond to the magnetic field components obtained from the Soloviev CME model, whereas the solid lines correspond to the mMT solution.}
    \label{fig:theoretical}
\end{figure*}

Before analyzing the thermodynamic and magnetic profiles obtained in EUHFORIA, it is insightful to examine how the components of the magnetic field evolve in the absence of solar wind. Figure~\ref{fig:theoretical} shows the evolution of the components of the magnetic field (in the Geocentric Solar Ecliptic System, GSE) for the two magnetic configurations at the level of three virtual satellites crossed by a torus with a speed of $600$ km/s. It is essential to note that the magnetic field depends on the geometry of the torus, notably through the aspect ratio: $R_{0}/a$. However, Fig.~\ref{fig:theoretical} provides a general idea of how the tilt of the torus and the in situ measurement point influence the components of the obtained magnetic fields. The minor and major radii of the torus are $a=5~\;R_{\odot}$ and $R=10~\;R_{\odot}$. For the Soloviev solution, the elongation, $\sigma$, equals 1, the triangularity, $\sigma$, is 0.5 and the parameter $\alpha_{S}$ also equals 1. The center of the torus evolves in the $X$ direction.

In the first row of Fig.~\ref{fig:theoretical}, the torus crosses the three virtual satellites without any tilt ($\omega=0$). When the virtual satellite is directly in the propagation path of the torus (i.e,\ the blue satellite in Fig.~\ref{fig:theoretical}) there are two sections corresponding to the front and back of the torus, separated by the torus hole. The two parts have the same size, and the distribution of the magnetic field inside depends on the CME model. In both cases, the $B_{x}$ component is zero, and the $B_{z}$ component is symmetric in the two sections, with a profile change sign close to the center. However, the direction of $B_{z}$ variation is opposite between the two CME models. Another difference is that the $B_{y}$ component, anti-symmetric between the two magnetic structures, is maximum close to the torus hole for the Soloviev solution, while it reaches its maximum at the magnetic axis for the mMT solution. 

If the virtual satellite is in the direction of propagation, it will pass through the torus hole. However, it is also possible to obtain a single magnetic structure by shifting a virtual satellite by an angle, $\phi=10^{\circ}$, so that it does not pass through the torus hole (cf.\ Fig.~\ref{fig:theoretical}, third and last column). As before, the two CME models have different magnetic field distributions, but a double change in the sign of the $B_{z}$ component is observed in both cases. A change in sign is also observed for the $B_{y}$ component. Moreover, unlike the case where the virtual satellite is in the direction of propagation, the $B_{x}$ component is not null. It has a significant influence and is even predominant near the structure's middle in the Soloviev solution (cf.\ the top-mid panel in Fig.~\ref{fig:theoretical}). Also, it is important to note that crossing either the right or left part of the torus (relative to the torus hole) merely changes the sign of $B_{x}$ in both models.

\citet{Asvestari22} found that the interaction of the CME with the ambient field in EUHFORIA can modify the inherent CME tilt even in the absence of actual CME rotation in the heliosphere. The middle and bottom rows of Fig.~\ref{fig:theoretical} show the magnetic field profiles in the two CME models obtained with a tilt, $\omega$, of $45^{\circ}$ and $-45^{\circ}$, respectively. The first observation is that at the three virtual satellites, there are two magnetic structures, indicating the passage of a part of the torus hole in all cases. However, the profiles are completely different from those obtained at the same location but without tilt. In particular, symmetry and anti-symmetry is no longer observed for the magnetic field components between the two parts of the torus when the virtual satellite is in the direction of propagation (cf.\ Fig.~\ref{fig:theoretical}, second column).

It should also be noted that the total force of the magnetic field (the black dashed and solid line in Fig.~\ref{fig:theoretical}) at the level of the three satellites is identical between the torus with a tilt of $\omega=45^{\circ}$ and the torus with an opposite tilt of $\omega=-45^{\circ}$. However, the distribution of the magnetic field is different, as are the dominant components. Even with tilt, the transition from a virtual satellite offset by an angle $\phi=10^{\circ}$ relative to the direction of propagation to a virtual satellite offset by an angle $\phi=-10^{\circ}$ only changes the sign of the component $B_{x}$.

Finally, in all configurations, the maximum reached by the total magnetic field is always higher in the Soloviev model than in the mMT model. In the latter, $B_{0}$ is the maximum at the magnetic axis, whereas in the Soloviev solution, $B$ can reach approximately $1.7\times B_{0}$ close to the boundary hole. However, this maximum depends on the parameter $\alpha_{S}$.

In conclusion, the magnetic field theoretically presents complex profiles that depend on the chosen model, its geometry, its tilt, and also the measurement point. Therefore, we can expect different profiles in EUHFORIA between the two models as well (cf.\ Sect.~\ref{sec:profileEUHFORIA}).

\subsection{Modelled profiles in EUHFORIA} \label{sec:profileEUHFORIA}

\subsubsection{Numerical set-up} \label{sec:setup}

For all the simulations we conducted with EUHFORIA for the present paper, the solar wind was reconstructed from the GONG magnetogram obtained at 01:04~UT on November 4, 2015. This date was chosen arbitrarily to illustrate the potential use of EUHFORIA. Hence, as mentioned in Sect.~\ref{sec:Introduction}, we are not trying to model the CMEs that occurred near this date, but we are interested in how the thermodynamic and magnetic profiles obtained during the propagation of the CME in the solar wind vary depending on the initial parameters.

The mesh used is uniform in all directions with an angular resolution of $4^\circ$ in the latitudinal and $2^\circ$ for the longitudinal directions and 256 cells in the radial direction. This resolution is similar to that used by \citet{Maharana22}. Convergence tests were carried out by doubling the number of cells in all three spatial directions, but this did not change our main results while drastically increasing the computation time.

Before studying the impact of different parameters, we carried out a series of simulations that will be used as a reference in the following sections. In these simulations, the center of the CME evolves from 12:04 PM on November 11 and from the initial position $O_{l}$ ($t_d=6.5\;R_{\odot}$, $\theta_{i}=81^{\circ}$, $\phi_{i}=0^{\circ}$) with a radial speed of $600\;$km/s. The torus has a minor radius $a=5\;R_{\odot}$, a major radius $\;R_{\odot}$, and a null tilt $\omega=0$. For the Soloviev solution, the triangularity is $\tau=0.5$ and the elongation $\sigma=1$. Thus, the geometric properties are identical to the tori presented in the previous section (cf.\ Sect.~\ref{sec:profileth}). We chose this elongation and triangularity because they allow a comparable cross section between the mMT and Soloviev models while maintaining a closed $\psi=1$ isocontour (cf.\ Sect.~\ref{sec:Soloviev_theory}).

At the points of intersection between the inner boundary of EUHFORIA and the CME (cf.\ Sect.~\ref{sec:implementation}), the injected temperature is constant at $0.8$~MK. The density is also constant at $n_{0}=1\times10^{-17}\;$kg\,m$^{-3}$. As explained in Sect.~\ref{sec:density}, using both a constant density and temperature results in a constant pressure within the torus. In fact, this contradicts the linear pressure profile we specified to obtain the Soloviev solution \citep{Soloviev75}. However, we have deliberately chosen to violate this assumption (cf.\ Sect.~\ref{sec:density} for more details) and use here only the magnetic distribution of the Soloviev equilibrium.

According to Fig.~\ref{fig:theoretical}, for the same value of $B_{0}$, the magnetic field strength is higher in the Soloviev solution than in the mMT model. Additionally, due to differing magnetic field distributions and slightly different geometries, the two CME models used do not have the same magnetic flux. Bearing this in mind, we still decided to use the same initial magnetic field strength, $B_{0}$, in both models, in order to align as closely as possible with the same initial parameters.

Finally, the simulations were run on the wICE cluster of the Vlaams Supercomputer Centrum, a Belgian institution specializing in high-performance computing\footnote{\url{http://www.vscentrum.be}}. They have utilized 4 nodes of this supercomputer, with 36 tasks per node, implying a total of 144 parallel processes. Additionally, each CPU allocated to this simulation has 2~GB of memory.

\subsubsection{References cases} \label{sec:reference}

\begin{figure}[ht!]
    \centering
    \includegraphics[width=0.5\textwidth]{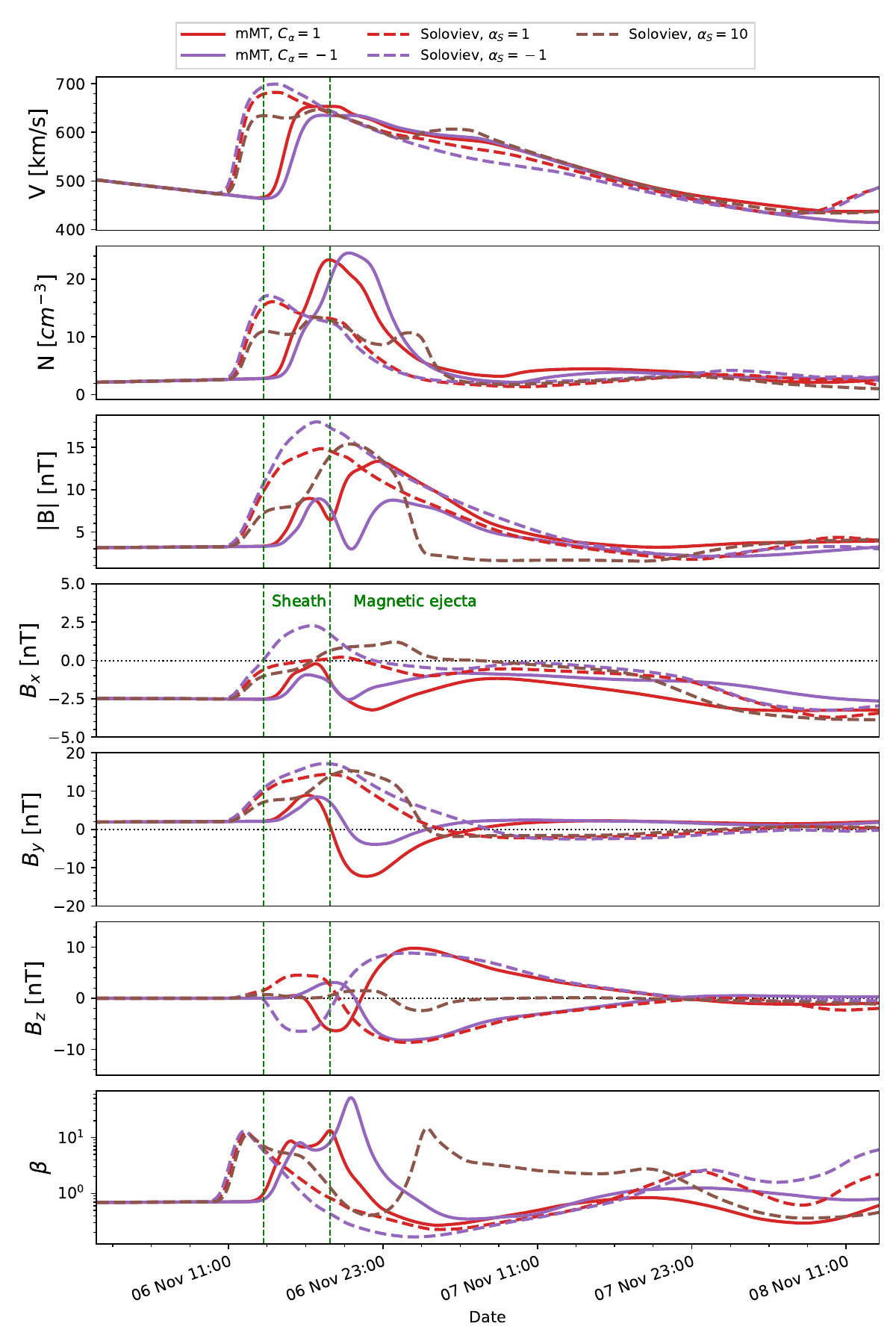}
    \caption{Magnetic and thermodynamic profiles as function of time obtained at Earth using EUHFORIA with the two CME models. Top to bottom : speed, density, total magnetic field strength, $B_{x}$, $B_{y}$, $B_{z}$ components in GSE coordinates, and plasma beta ($\beta$ in log scale). The solid lines correspond to simulations where the CME is based on the mMT solution, while the dashed lines are obtained from simulations with the Soloviev solution as CME model. The green vertical lines mark the presumed sheath boundaries in the mMT simulation with $C_{\alpha}=1$.}
    \label{fig:reference}
\end{figure} 

\begin{figure*}[h!]
    \begin{subfigure}{0.49\linewidth} 
    \centering\includegraphics[width=\linewidth]{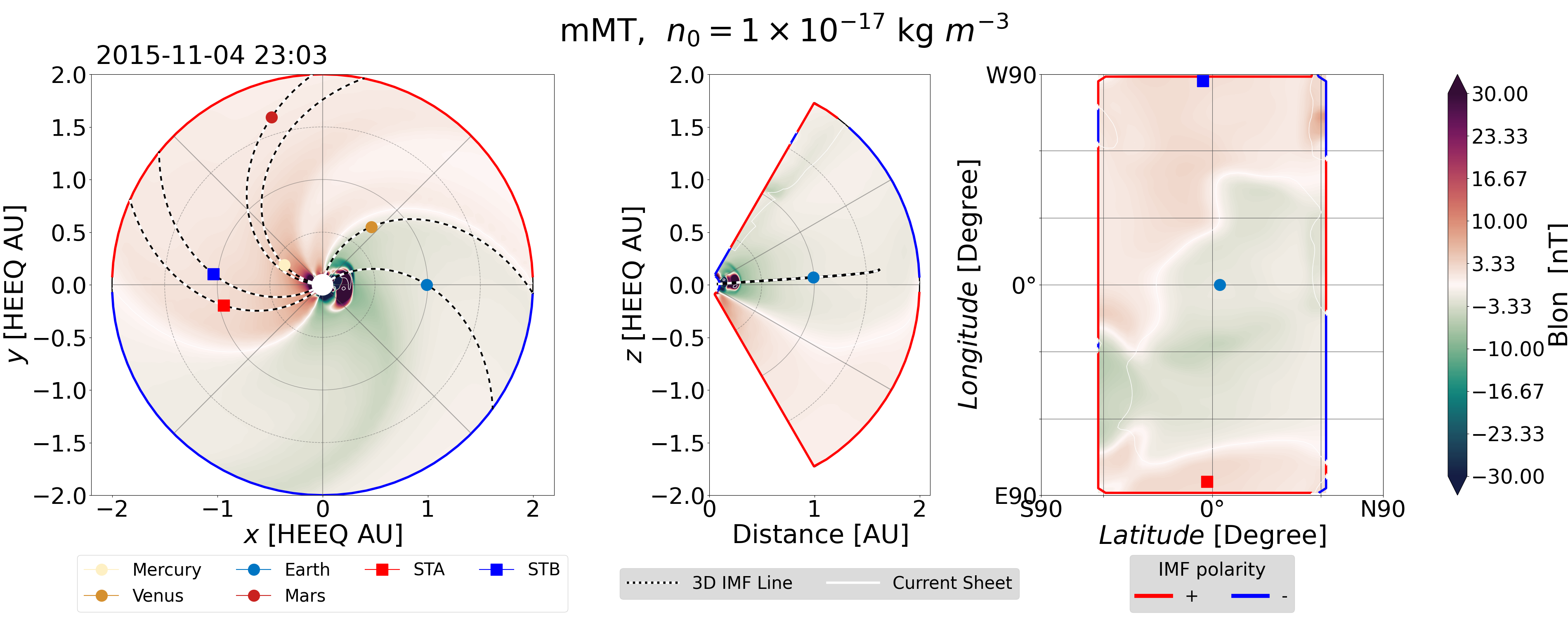}
    \caption{mMT CME with $n_{0}=1\times 10^{-17}\;$kg\,m$^{-3}$; 4 November 2015 at 23:02} \label{fig:blona}
    \end{subfigure}
    \hfill
    \begin{subfigure}{0.49\linewidth}
    \centering\includegraphics[width=\linewidth]{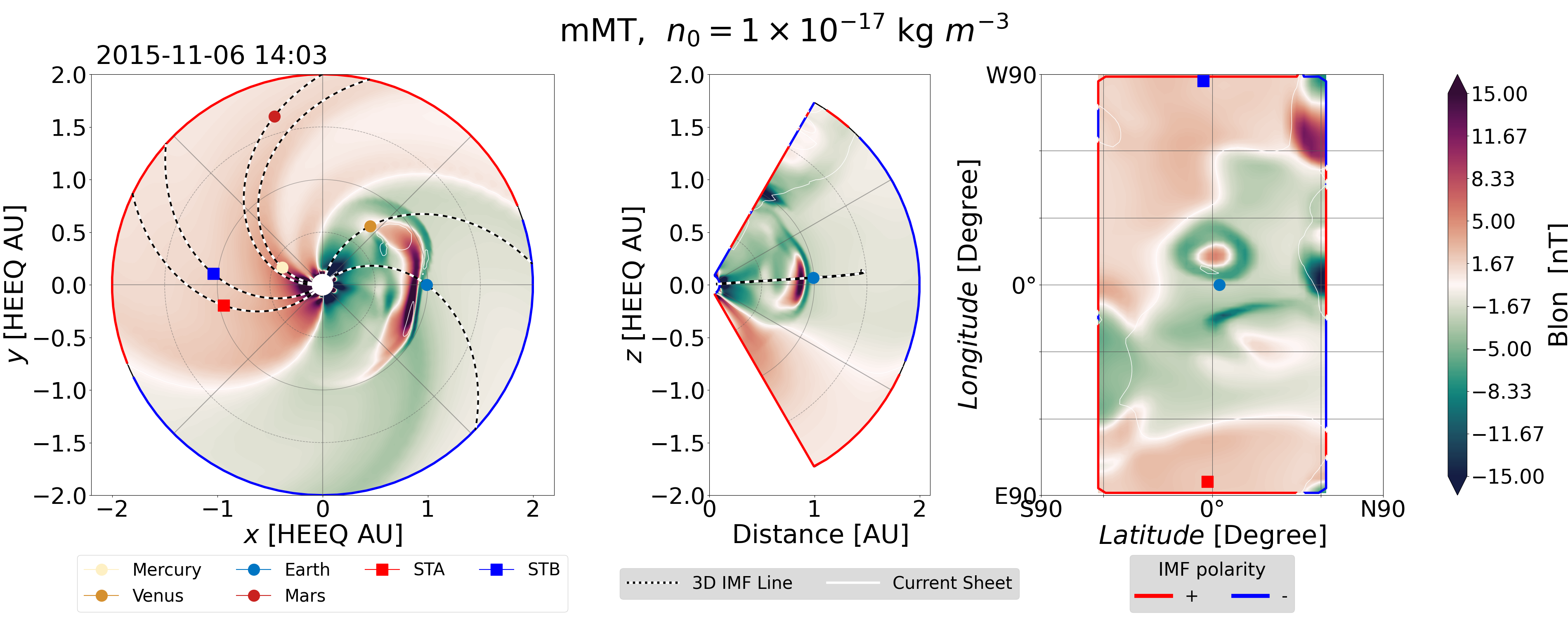}
    \caption{mMT CME with $n_{0}=1\times 10^{-17}\;$kg\,m$^{-3}$; 6 November 2015 at 14:03} \label{fig:blonb}
    \end{subfigure}
    \par\bigskip
    \begin{subfigure}{0.49\linewidth} 
    \centering\includegraphics[width=\linewidth]{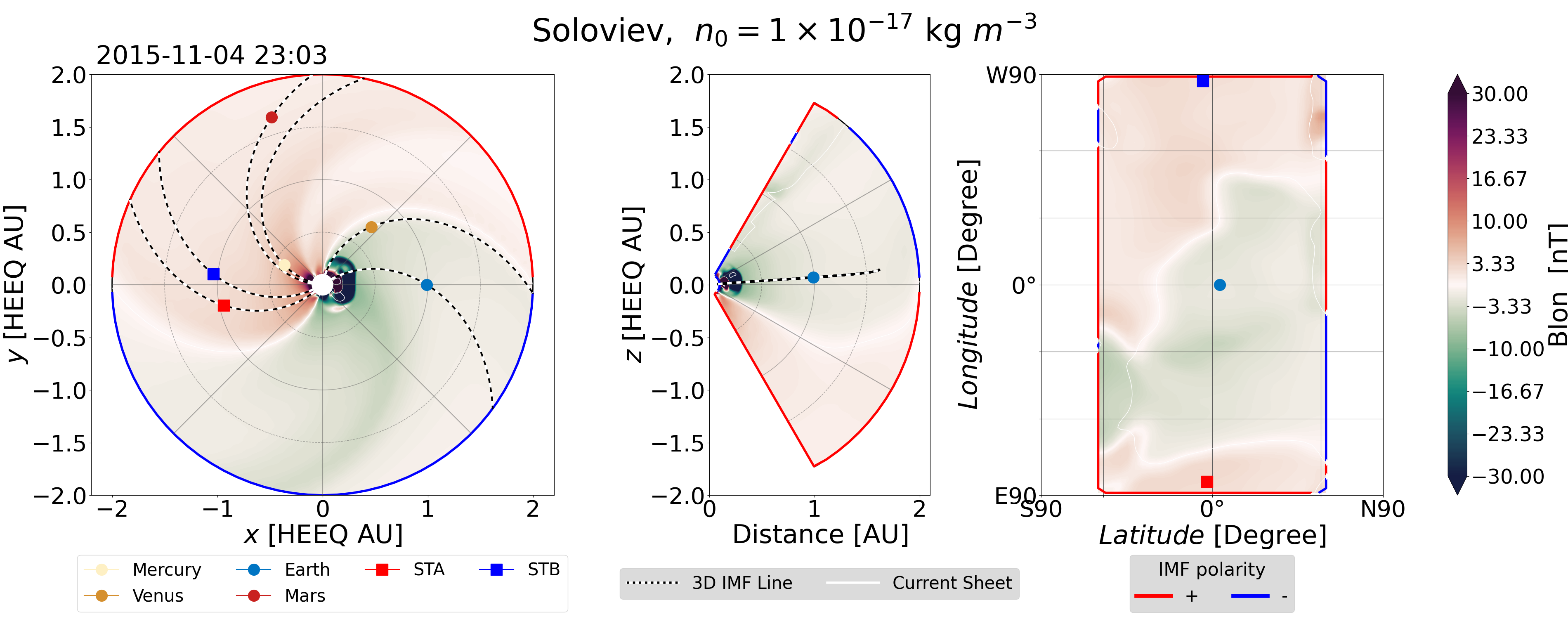}
    \caption{Soloviev CME with $n_{0}=1\times 10^{-17}\;$kg\,m$^{-3}$; 4 November 2015 at 23:03}\label{fig:blonc}
    \end{subfigure}
    \hfill
    \begin{subfigure}{0.49\linewidth}
    \centering\includegraphics[width=\linewidth]{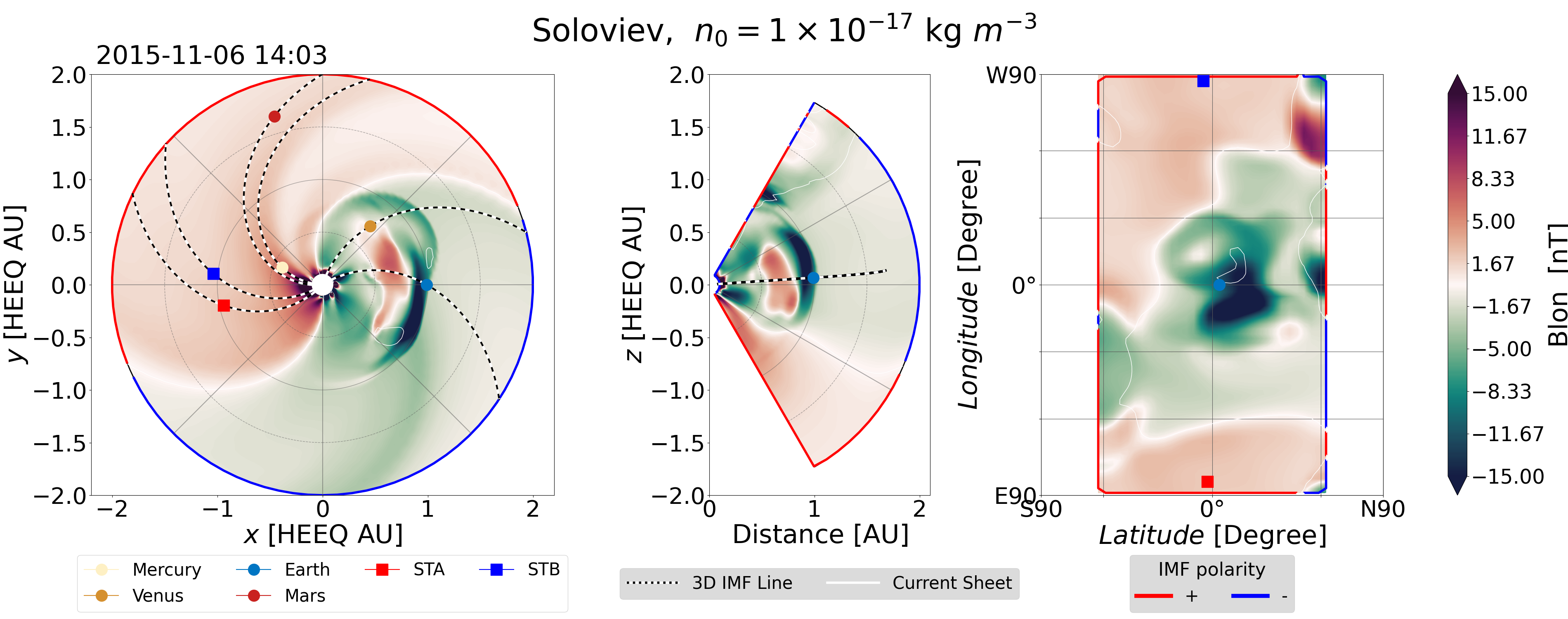}
    \caption{Soloviev CME with $n_{0}=1\times 10^{-17}\;$kg\,m$^{-3}$; 6 November 2015 at 14:03}\label{fig:blond}
    \end{subfigure}
    \par\bigskip
    \begin{subfigure}{0.49\linewidth} 
    \centering\includegraphics[width=\linewidth]{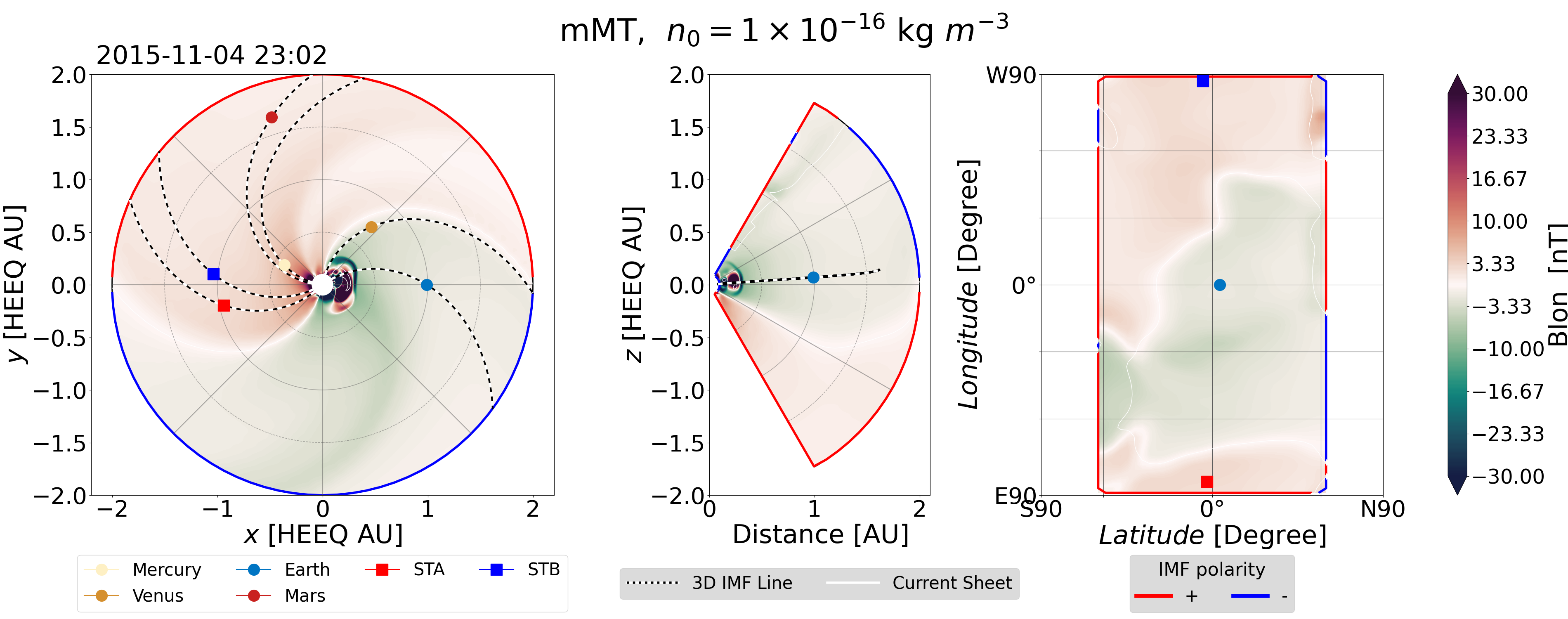}
    \caption{mMT CME with $n_{0}=1\times 10^{-16}\;$kg\,m$^{-3}$; 4 November 2015 at 23:02}\label{fig:blone}
    \end{subfigure}
    \hfill
    \begin{subfigure}{0.49\linewidth}
    \centering\includegraphics[width=\linewidth]{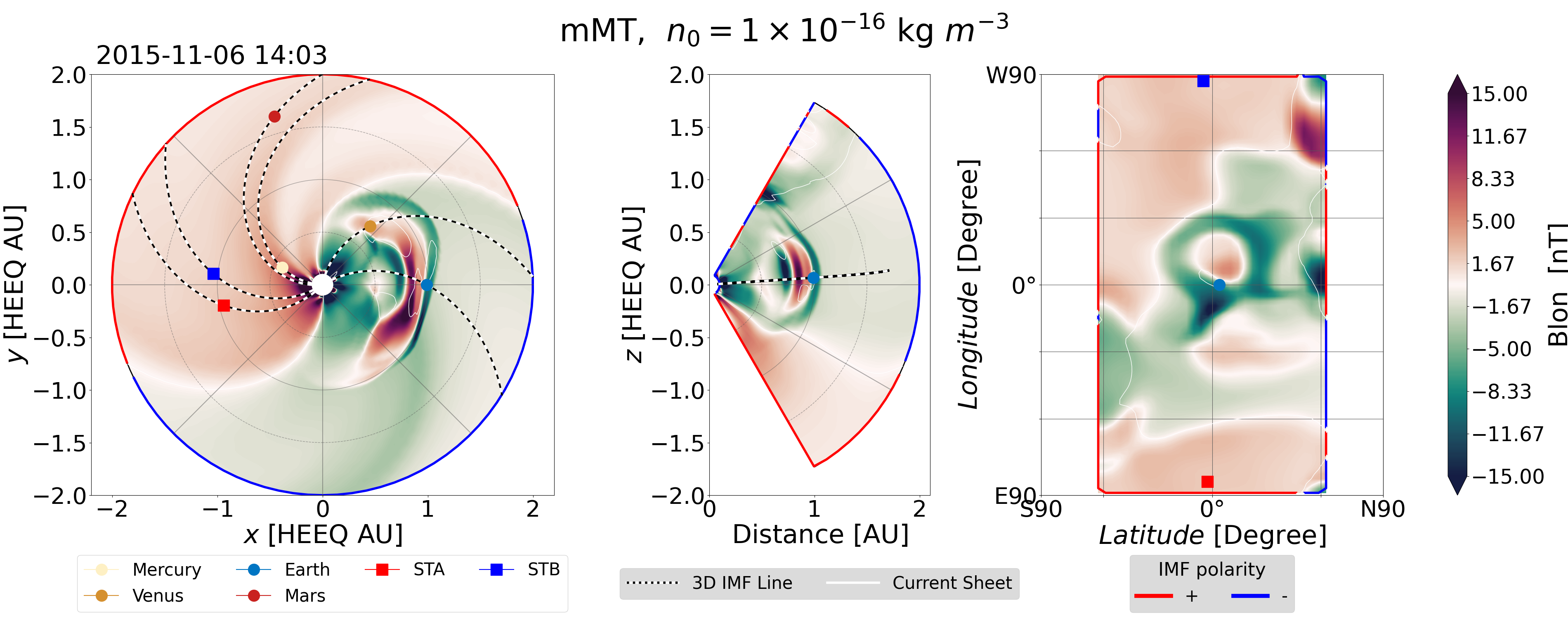}
    \caption{mMT CME with $n_{0}=1\times 10^{-16}\;$kg\,m$^{-3}$; 6 November 2015 at 14:03}\label{fig:blonf}
    \end{subfigure}
    \par\bigskip
    \begin{subfigure}{0.49\linewidth} 
    \centering\includegraphics[width=\linewidth]{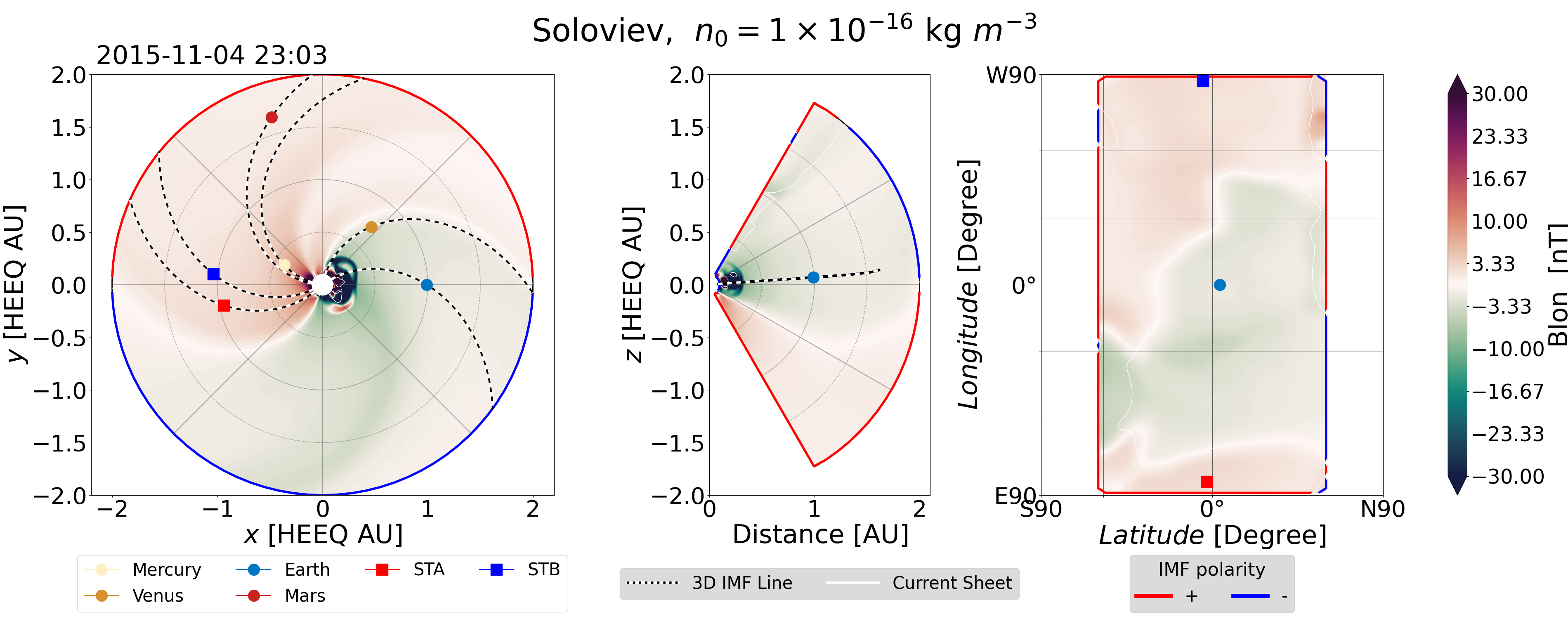}
    \caption{Soloviev CME with $n_{0}=1\times 10^{-16}\;$kg\,m$^{-3}$; 4 November 2015 at 23:03}\label{fig:blong}
    \end{subfigure}
    \hfill
    \begin{subfigure}{0.49\linewidth}
    \centering\includegraphics[width=\linewidth]{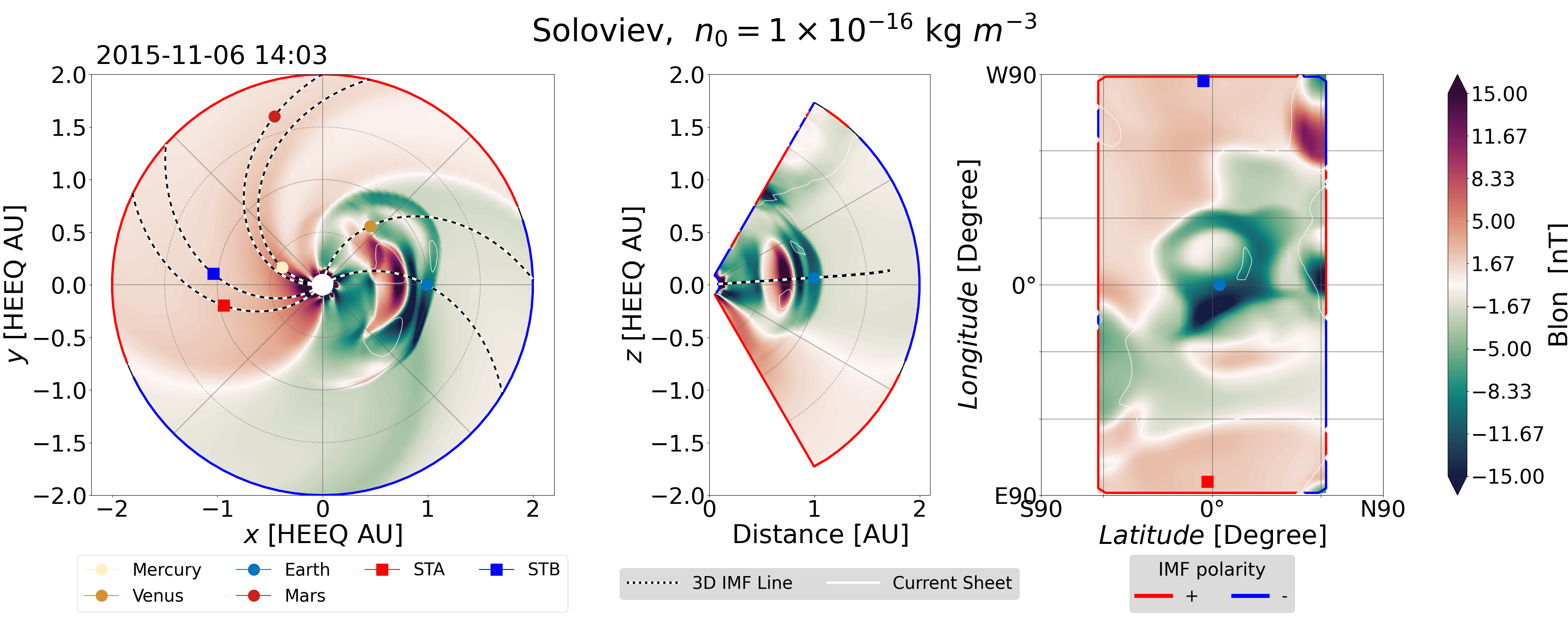}
    \caption{Soloviev CME with $n_{0}=1\times 10^{-16}\;$kg\,m$^{-3}$; 6 November 2015 at 14:03}\label{fig:blonh}
    \end{subfigure}
    \caption{Visualization of the longitudinal magnetic field component in the EUHFORIA heliosphere domain. For both CME models, two initial mass densities were used: $n_{0}=1\times 10^{-16}\;$kg\,m$^{-3}$ (panels a and b for the mMT CME model and panels c and d for the Soloviev CME model) and $n_{0}=1\times 10^{-16}\;$kg\,m$^{-3}$ (panels e and f for the mMT CME model and panels g and h for the Soloviev CME model). The geometrical, magnetic, and thermodynamic parameters of the CMEs are described in Sect.~\ref{sec:setup}. The left panels (a, c, e, g) show the evolution of the longitudinal field, $B_{lon}$, on March 4 while the right panels (b, d, f, h) show the simulation on November 6. Each panel is divided into an equatorial cross section, a meridional cross section passing through the Sun and the Earth, and a Mercator projection at 0.1~AU from the Sun. The visualization displays the positions of the different planets in the simulation domain, the location of the current sheet, and the interplanetary magnetic field connecting the Sun and the planets, obtained from Parker spirals.} \label{fig:Blon}
\end{figure*}

According to the initial magnetic field and the geometrical parameters outlined earlier, Fig.~\ref{fig:reference} illustrates the evolution of speed, density, magnetic field, and plasma beta ($\beta$) at Earth in EUHFORIA for various CME configurations. We simulated three CMEs using the Soloviev solution with differing $\alpha_{S}$ ($\alpha_{S}=1$, $-1$, and $10$). Additionally, we modeled two CMEs with the mMT solution with $C_{\alpha}=1$ and $-1$, respectively.

In Fig.~\ref{fig:reference}, we observe that using either the mMT or the Soloviev solution as the CME model results in notably different thermodynamic and magnetic profiles. Specifically, although the CMEs have the same initial speed and the same initial magnetic field strength, $B_{0}$, the speed obtained at Earth is higher in simulations with a Soloviev CME than with a mMT CME. For example, the CME computed from the Soloviev model with $\alpha_{S}=1$ reaches a maximum speed that is approximately $9\%$ higher than the mMT CME with $C_{\alpha}=1$. Consequently, the shock arrival time is dependent on the chosen CME model. Additionally, the Soloviev CMEs also demonstrate a higher magnetic field strength, $|B|$, than the mMT CMEs, as anticipated from the theoretical profiles (cf.\ Fig.~\ref{fig:theoretical}). This suggests that the speed of the CME is not solely related to the initially imposed speed but also to the initial magnetic field strength (cf.\ Sect.~\ref{sec:speed}).

At the very end of the simulation, after November 8 in Fig.~\ref{fig:reference}, the speed increases in the two fastest simulations. This increase is a numerical artifact arising from the solver's challenges in managing steep speed gradients as detailed in Sect.~4.2 of \citet{Linan23}. This can be partially mitigated by enhancing the spatial resolution.

\begin{figure*}[ht!]
    \begin{subfigure}{0.49\linewidth} 
    \centering\includegraphics[width=\linewidth]{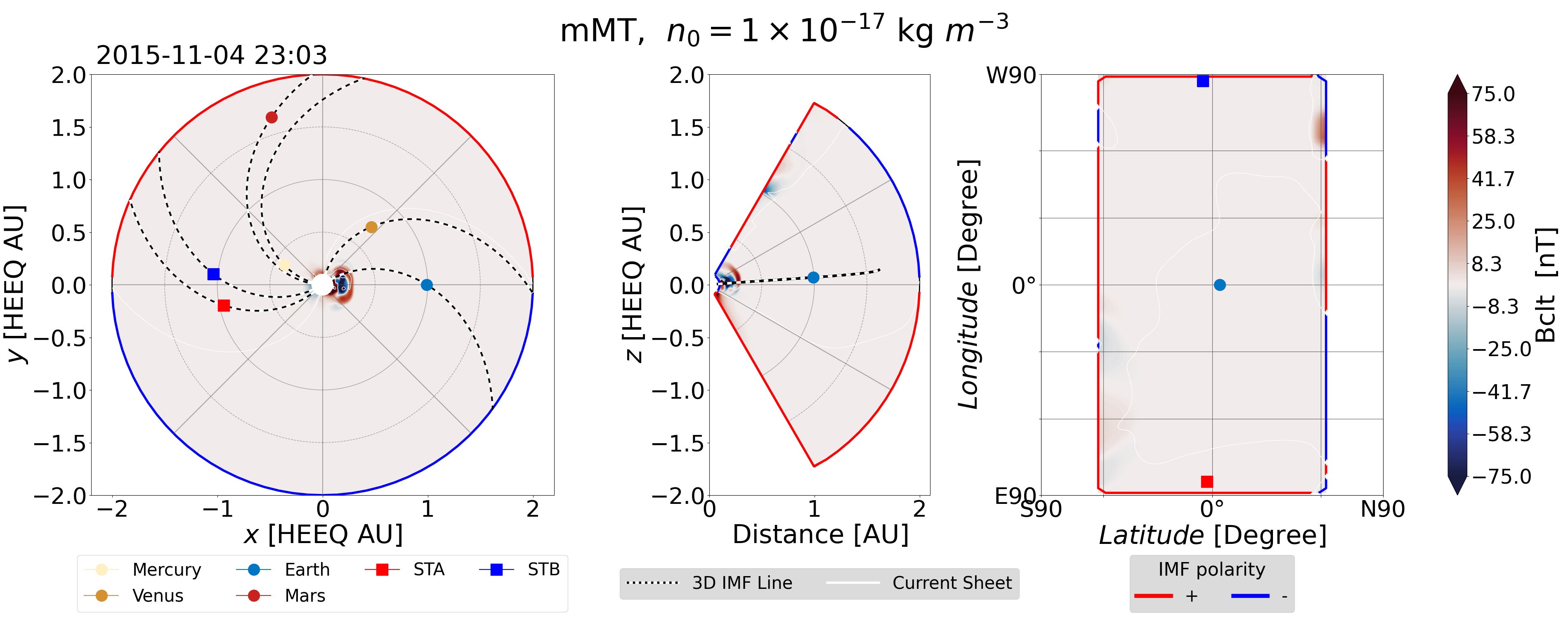}
    \caption{mMT CME with $n_{0}=1\times 10^{-17}\;$kg\,m$^{-3}$; 4 November 2015 at 23:02} \label{fig:bclta}
    \end{subfigure}
    \hfill
    \begin{subfigure}{0.49\linewidth}
    \centering\includegraphics[width=\linewidth]{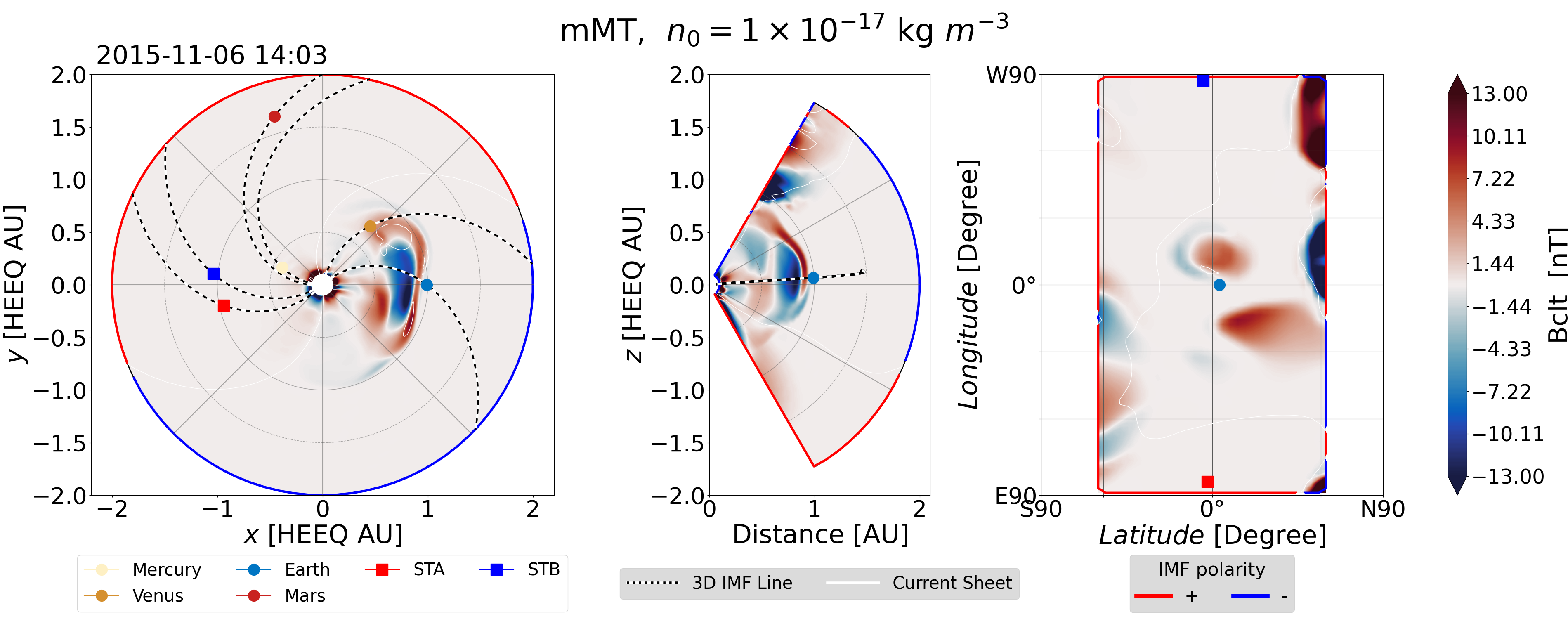}
    \caption{mMT CME with $n_{0}=1\times 10^{-17}\;$kg\,m$^{-3}$; 6 November 2015 at 14:03} \label{fig:bcltb}
    \end{subfigure}
    \par\bigskip
    \begin{subfigure}{0.49\linewidth} 
    \centering\includegraphics[width=\linewidth]{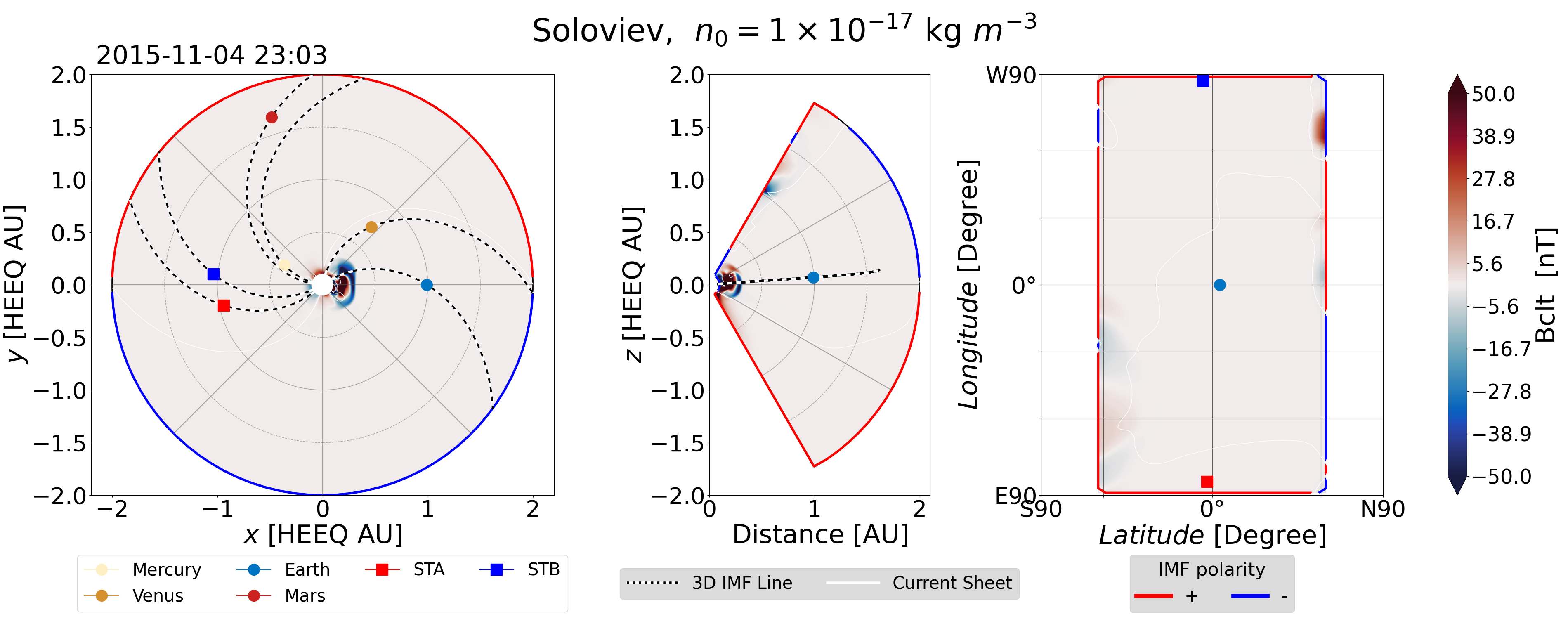}
    \caption{Soloviev CME with $n_{0}=1\times 10^{-17}\;$kg\,m$^{-3}$; 4 November 2015 at 23:03}\label{fig:bcltc}
    \end{subfigure}
    \hfill
    \begin{subfigure}{0.49\linewidth}
    \centering\includegraphics[width=\linewidth]{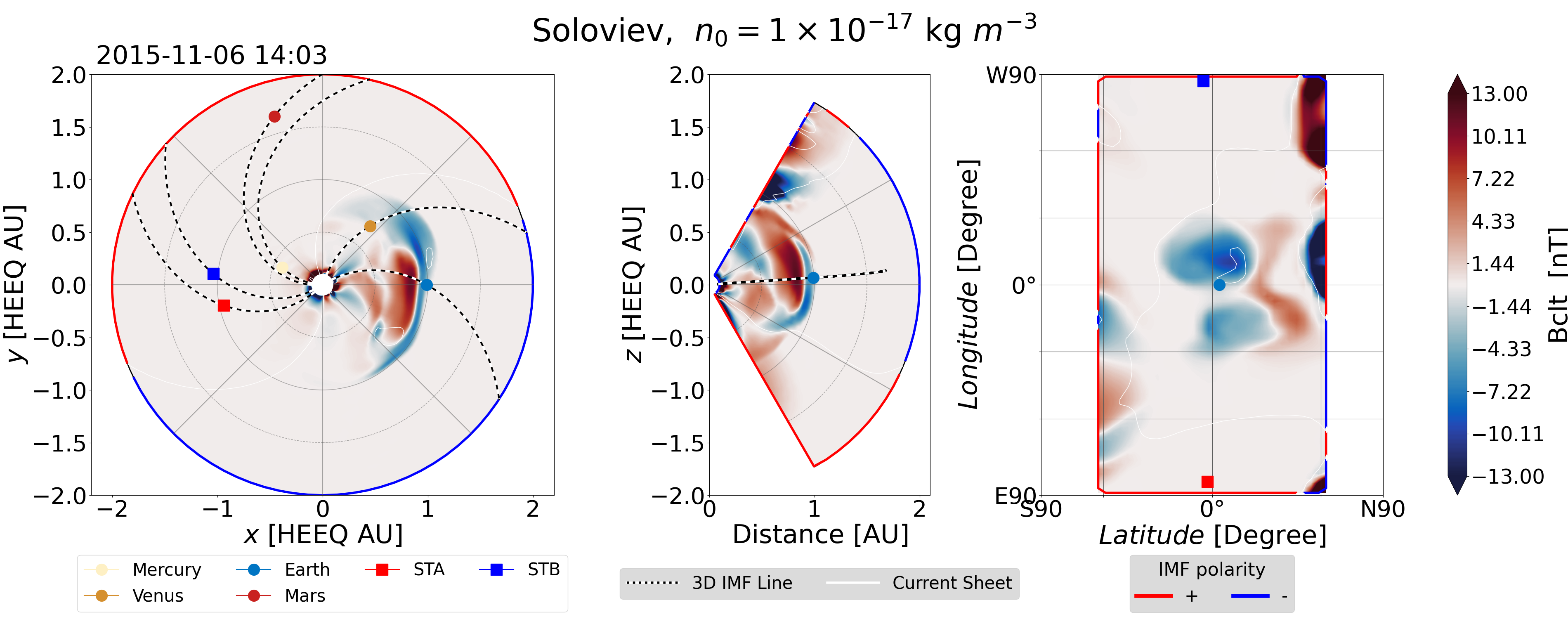}
    \caption{Soloviev CME with $n_{0}=1\times 10^{-17}\;$kg\,m$^{-3}$; 6 November 2015 at 14:03}\label{fig:bcltd}
    \end{subfigure}
    \par\bigskip
    \begin{subfigure}{0.49\linewidth} 
    \centering\includegraphics[width=\linewidth]{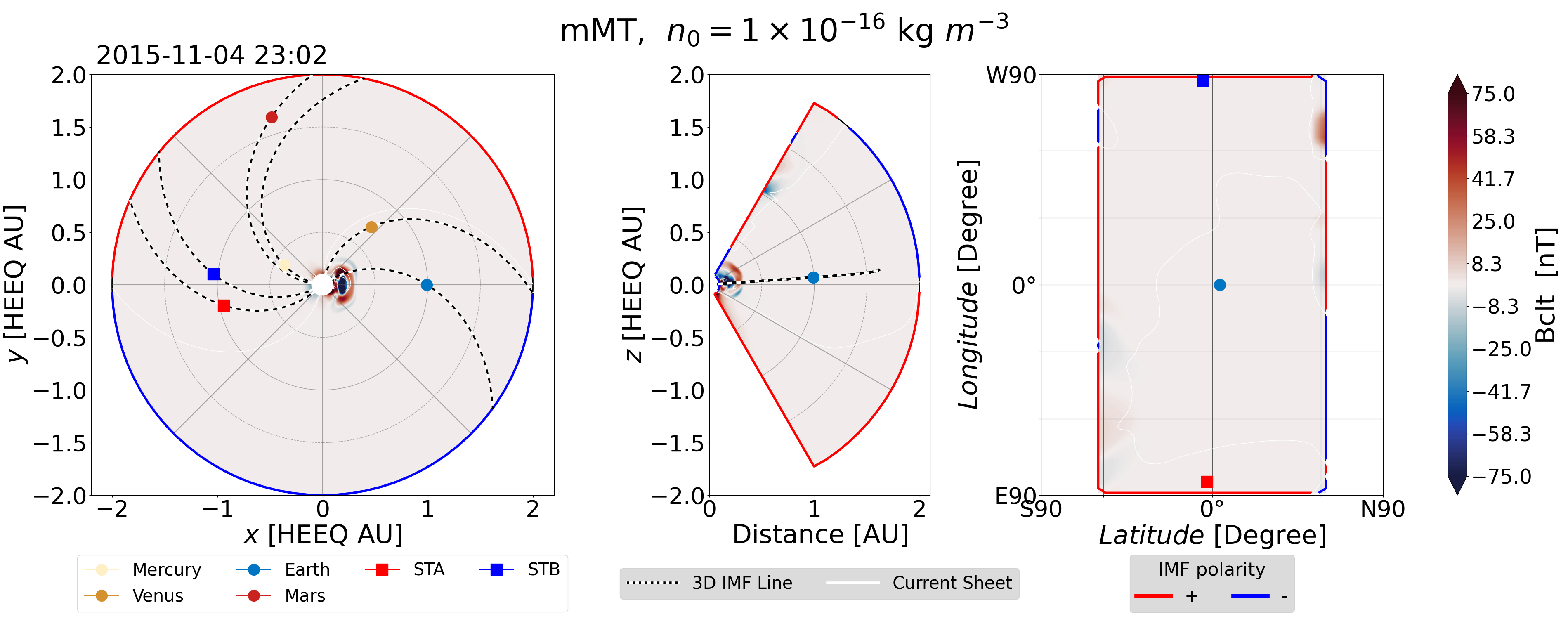}
    \caption{mMT CME with $n_{0}=1\times 10^{-16}\;$kg\,m$^{-3}$; 4 November 2015 at 23:02}\label{fig:bclte}
    \end{subfigure}
    \hfill
    \begin{subfigure}{0.49\linewidth}
    \centering\includegraphics[width=\linewidth]{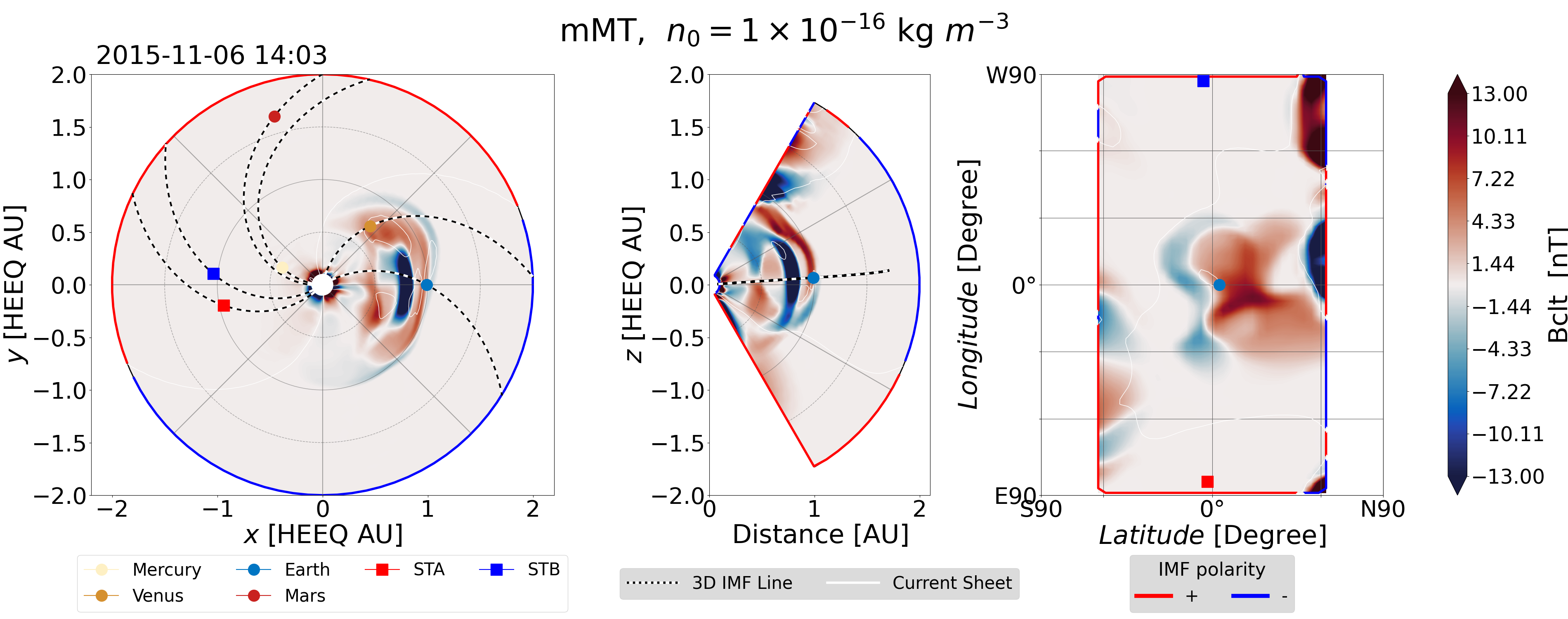}
    \caption{mMT CME with $n_{0}=1\times 10^{-16}\;$kg\,m$^{-3}$; 6 November 2015 at 14:03}\label{fig:bcltf}
    \end{subfigure}
    \par\bigskip
    \begin{subfigure}{0.49\linewidth} 
    \centering\includegraphics[width=\linewidth]{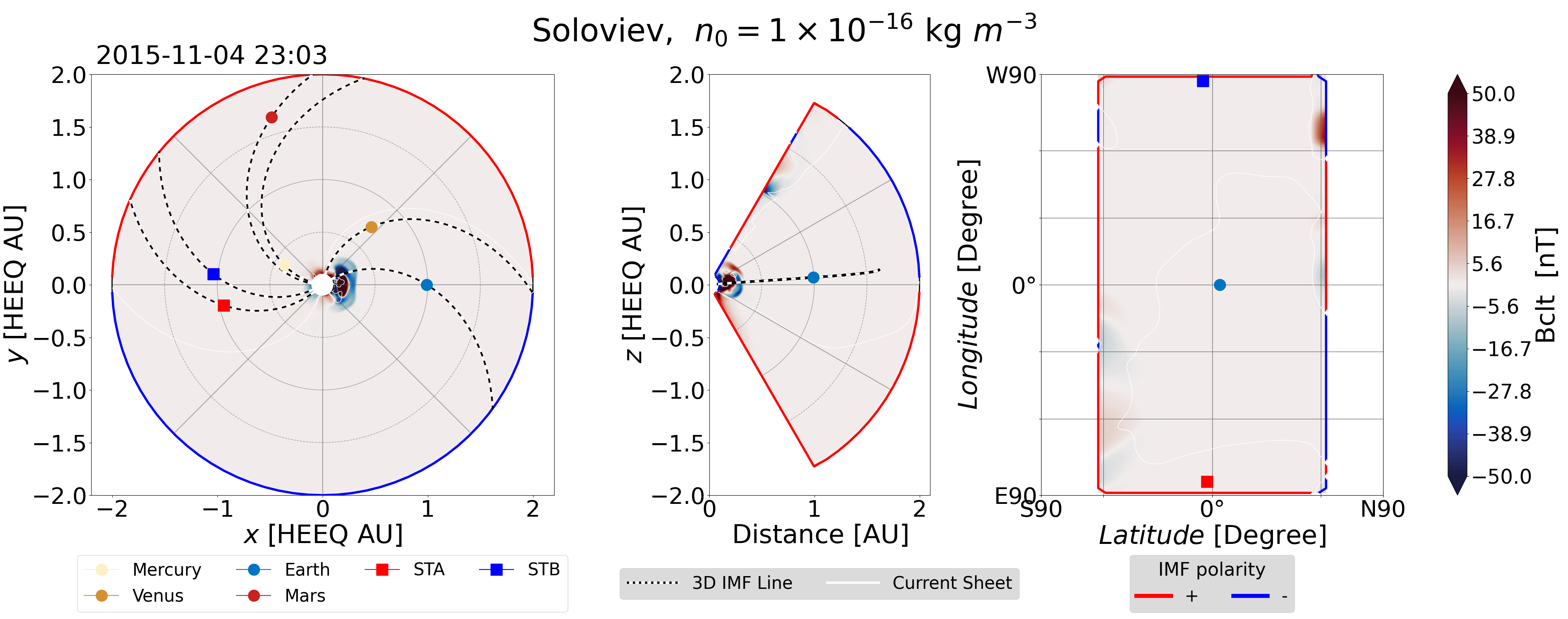}
    \caption{Soloviev CME with $n_{0}=1\times 10^{-16}\;$kg\,m$^{-3}$; 4 November 2015 at 23:03}\label{fig:bcltg}
    \end{subfigure}
    \hfill
    \begin{subfigure}{0.49\linewidth}
    \centering\includegraphics[width=\linewidth]{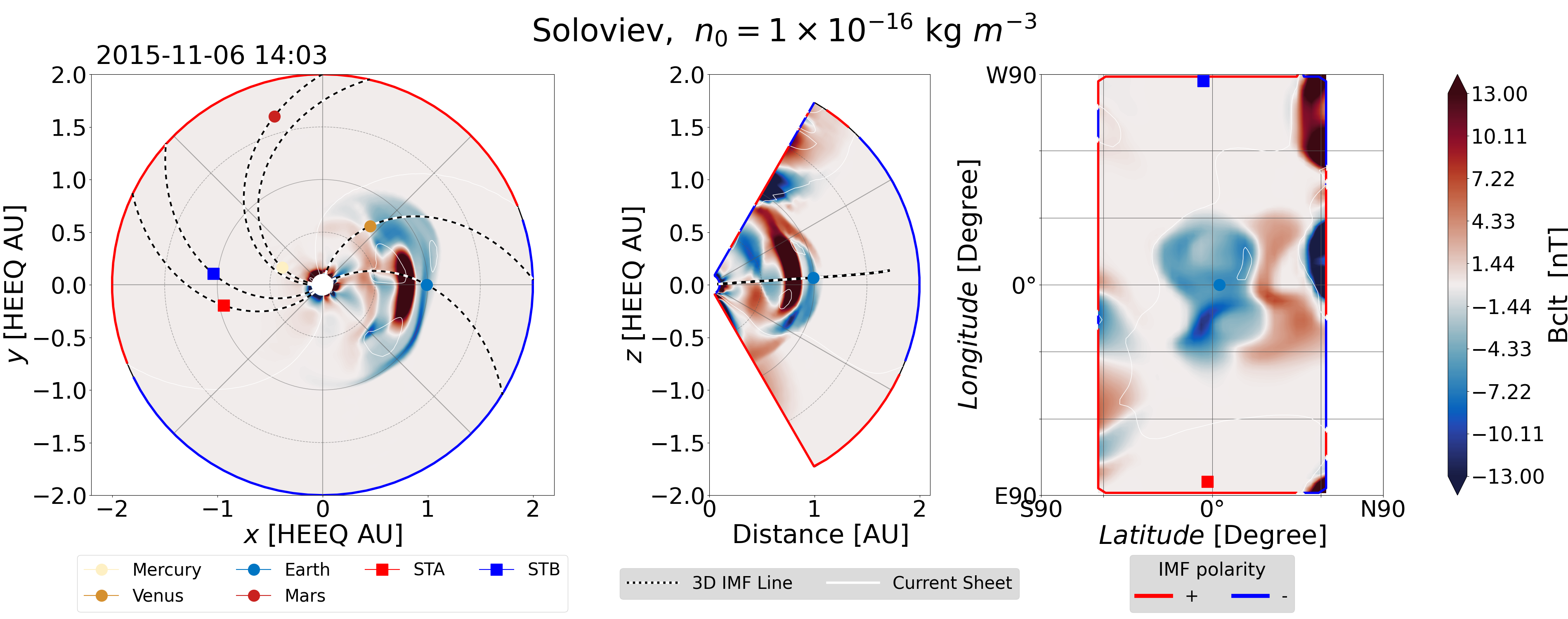}
    \caption{Soloviev CME with $n_{0}=1\times 10^{-16}\;$kg\,m$^{-3}$; 6 November 2015 at 14:03}\label{fig:bclth}
    \end{subfigure}
    \caption{Same as Fig.~\ref{fig:Blon} but for the co-latitudinal magnetic field, $B_{clt}$.} \label{fig:Bclt}
\end{figure*}

As discussed in Sect.~\ref{sec:density}, even if the CMEs are inserted with the same density, the peak density measured at Earth differs in various simulations. The temporal evolution of the parameter $\beta$ also depends on the CME model used. However, for all CMEs, an increase in the parameter $\beta$ is observed at the moment of arrival at Earth, followed by a decrease to values lower than those of the solar wind before the event. The Soloviev CME with $\alpha_{S}=10$ exhibits a distinct evolution from other cases, as there is a second increase in the parameter $\beta$ a few hours after the first increase, which might indicate the passage of the back part of the CME or some perturbation in the wake of the CME. Except for this case, the profile of $\beta$ is consistent with real in situ measurements on Earth. Indeed, \citet{Regnault20}, by performing superposed epoch analyzes of 20 years of ACE data, found that during an active period of the solar cycle, the parameter $\beta$ increases in the sheath but is lower than before the event during the traversal of the magnetic ejecta.

Considering the magnetic field profiles in Fig.~\ref{fig:reference}, the results align with the theoretical profiles presented in Sect.~\ref{fig:theoretical}, suggesting that they are influenced by the specific CME model implemented. All CME models feature a non-zero $B_{x}$ component, implying that Earth is not perfectly in the CME propagation direction (cf.\ Fig.~\ref{fig:theoretical}). However, its contribution is minor compared to the $B_{y}$ and $B_{z}$ components, whose trends vary from one CME to another.

To explain the trend of the $B_{y}$ component, Fig.~\ref{fig:Blon} shows the distribution of the longitudinal magnetic field $B_{lon}$ at two different times for the CME derived from the mMT solution with positive chirality (Figs.~\ref{fig:blona} and \ref{fig:blonb}) and from the Soloviev solution (Figs.~\ref{fig:blonc} and \ref{fig:blond}). The longitudinal magnetic field $B_{lon}$ in the EUHFORIA coordinate system is directly related to the $B_{y}$ component in the geocentric solar ecliptic system ($B_{y}=-B_{lon}$). Initially (cf.\ Fig.~\ref{fig:blona}), $B_{lon}$ is positive at the front part of the CME and negative at the back in the mMT model. For the Soloviev CME model with $\alpha=1$ (cf.\ Fig.~\ref{fig:blonc}), the $B_{lon}$ distribution is almost similar but with opposite signs, as anticipated by the theoretical profiles. We can also note that the longitudinal field in the torus hole is slightly positive in the mMT solution and negative in the Soloviev solution. The magnetic field within the torus closely resembles that of the CME front part. We suggest that it is a consequence of the solver ensuring the magnetic field's solenoidality and absence of significant gradients.

Upon the arrival of the magnetic ejecta at Earth (Figs.~\ref{fig:blonb} and \ref{fig:blond}), the torus expanded in all directions following a self-similar expansion. The initial toroidal geometry is no longer discernible. Furthermore, ahead of the front part of the CME derived from the mMT model, we can see a thin region where the longitudinal field is negative. Since the longitudinal field of the CME is positive in the front part of the CME (cf.\ Fig.~\ref{fig:theoretical}), we associate this negative field with the presence of a sheath that develops as the CME expands \citep{Siscoe08}.

With respect to the different time profiles, we suggest a transition between the sheath and the magnetic ejecta at approximately 2015-11-06T19:00, (cf.\ the vertical green line in Fig.~\ref{fig:reference}) for the mMT simulation with $C_{\alpha}=1$ (the full red curve). In this simulation, the passage through the sheath is characterized by an increase in the $B_{y}$-component. On the contrary, the passage through the magnetic ejecta is marked by an extended duration, where $B_{y}$ remains negative. These two distinct structures result in two main peaks in the total magnetic field strength, $|B|$. However, for the Soloviev CME cases, we observe mainly an increase in both the $B_{y}$-component and $|B|$. Since the longitudinal field in the front part of the CME is negative, there is no sign change in the $B_{y}$-component in this case. Consequently, the presence of a sheath is not as evident as in the mMT CME. We suggest that the rise in plasma $\beta$ could indicate the passage through a thin sheath \citep{Masias16}. However, solely based on the thermodynamic and magnetic profiles shown in Fig.~\ref{fig:reference}, it is not feasible to pinpoint the exact time of the transition between the sheath and the magnetic ejecta in the Soloviev simulations. It is also worth noting that determining this boundary can be complicated due to magnetic reconnection occurring at various points along the ME, which smooth the transition between the different magnetic structures, as discussed by \citep{Romashets03}. Moreover, the sheath forms due to turbulence which is not included in the simulation model. Further studies beyond the scope of this work are thus necessary to gain a better understanding of the development and characteristics of sheaths in EUHFORIA. The primary concern is that in EUHFORIA, the sheath only develops after the CME injection at 0.1~AU. By not considering the interactions between the solar wind and the CME closer to the Sun, the size and properties of the sheath can deviate significantly from expectations, potentially leading to inconsistencies when compared to in-situ measurements at L1.

Finally, based on the theoretical profiles (cf.\ Fig.~\ref{fig:theoretical}), we expected to see a significant variation of the $B_{y}$ field at the simulation end, indicating the passage of the torus's back. In theory, this contribution should be antisymmetric with respect to the torus front (i.e.,\ having the same amplitude but with the opposite sign). The back of the torus, characterized by a positive longitudinal field in the Soloviev case, is visible in Fig.~\ref{fig:blond}. However, its amplitude is so weak that it is not detectable in the time evolution of the component $B_{y}$ in Fig.~\ref{fig:reference}.

Figure~\ref{fig:Bclt} shows the evolution of the colatitudinal magnetic field, $B_{clt}$, for the same two CMEs. In the EUHFORIA coordinate system, $B_{clt}$ equals $-B_{z}$ in GSE. Initially (cf.\ Figs.~\ref{fig:bclta} and \ref{fig:bcltc}), we distinguish the injected torus through the inner boundary, with a change in the sign of the $B_{z}$ component as predicted by the theoretical profiles (cf.\ Fig.~\ref{fig:theoretical}). Like $B_{lon}$, the magnetic field within the torus hole follows the trend of the last point of the CME injected before it. Later in the propagation (cf.\ Figs.~\ref{fig:bcltb} and \ref{fig:bcltd}), the inversion line in the front part is clearly visible in both cases. As a consequence of this magnetic field distribution, the time evolution of the $B_{z}$ component presents a significant change in sign. However, as with the $B_{y}$ component, the temporal profile of $B_{z}$ does not allow us to distinguish the influence of the back part of the CME (cf.\ Fig.~\ref{fig:reference}). Further analysis is necessary to understand the exact causes of erosion and/or dispersion of the back part.

Theoretically, the sign of the $B_{z}$ component is controlled by the sign of chirality, i.e.,\ the sign of $C_{\alpha}$ or $\alpha_{S}$. In EUHFORIA, changing the sign of chirality indeed reverses the profile of $B_{z}$ (cf.\ Fig.~\ref{fig:reference}). However, we also notice that chirality influences the amplitude of the different components, particularly in the simulations with an mMT CME. For example, the maximum of the total magnetic field strength, $|B|$, is $13.4$ nt for the mMT CME with $C_{\alpha}=1$, while $|B|$ reaches only $9.0$ nt for the mMT CME with $C_{\alpha}=-1$. This leads us to infer that the distribution of the magnetic field affects the nature of the interactions with the ambient solar wind, thus influencing the predicted profiles.

Finally, as expected, increasing the $\alpha_{S}$ parameter in the simulations with the Soloviev solution allows for a more longitudinal field. Therefore, in the case of the CME with $\alpha_{S}=10$, the predominant magnetic contribution is the $B_{y}$ component. In the forthcoming sections, we will focus exclusively on CMEs with $\alpha_{S}=1$ to facilitate comparisons with CMEs derived from the mMT solution, which show a substantial contribution from the $B_{z}$ component.

\subsection{Impact of the initial velocity} \label{sec:speed}

\begin{figure}[h]
    \centering
    \includegraphics[width=0.5\textwidth]{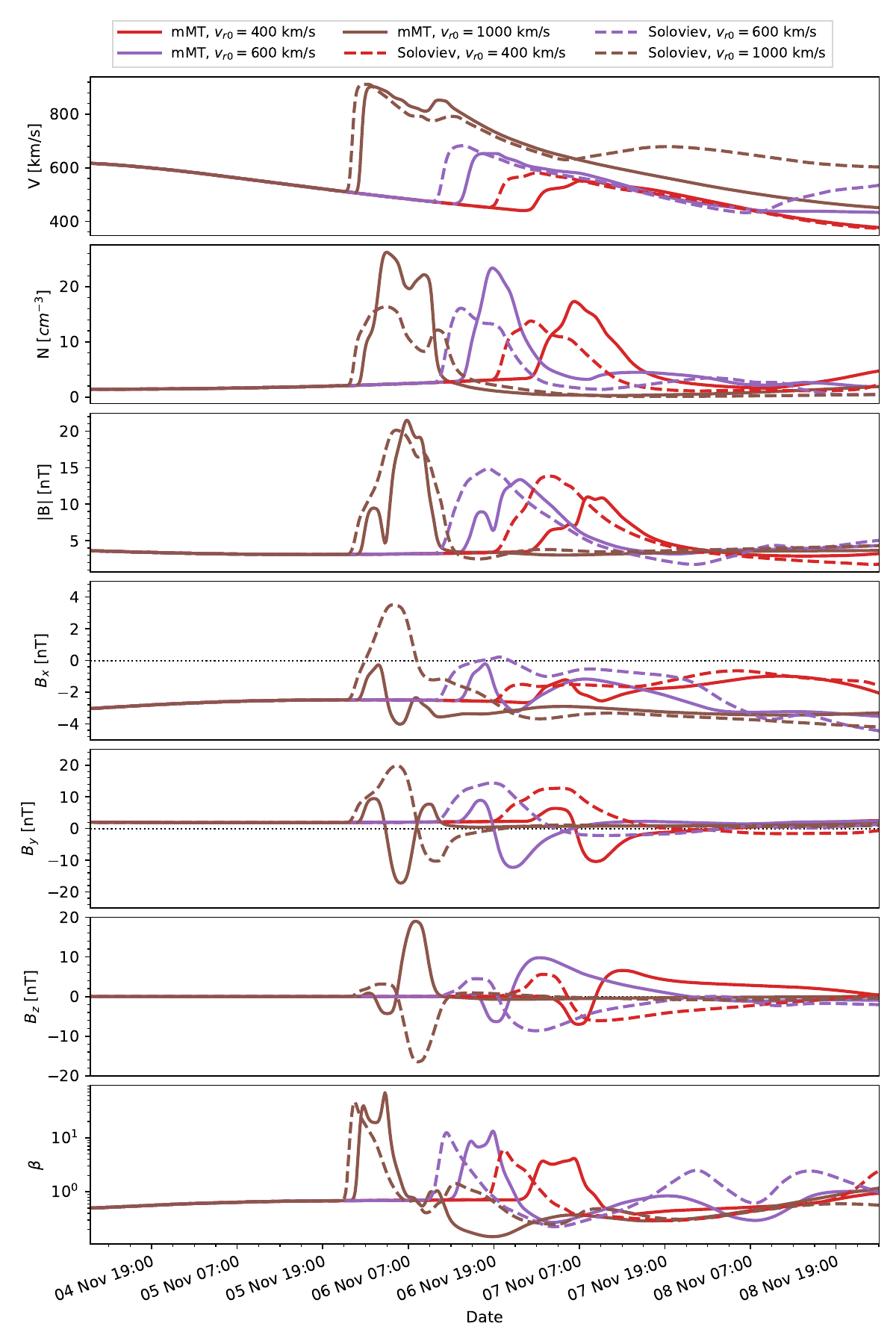}
    \caption{Same as Fig.~\ref{fig:reference}. The only difference between the different cases computed from the same CME model is the initial radial speed. For both CME solution, the initial speed is $v_{r0}=300$ km/s, $v_{r0}=600$ km/s and $v_{r0}=900$ km/s.}
    \label{fig:speed}
\end{figure}

As outlined in Sect.~\ref{sec:implementation}, the center of the torus moves in a purely radial direction at each time step, according to the following equation:
\begin{equation}
t_{d}(t)=t_{d}(0)+v_{r0} t
\end{equation}
where $t_{d}(t)$ denotes the distance from the domain's origin at time $t$, and $v_{r0}$ represents the initial radial speed. Consequently, the torus experiences a translation in the direction of vector $OO_{l}$ (cf.\ Fig.~\ref{fig:system}).

At EUHFORIA's inner boundary, the solar wind speed is altered to model the CME's passage. There are two primary strategies to set the injected velocity. The first strategy is to compute the velocity at each point on the torus cross-section, considering its translation and the imposed radial velocity at the center. Therefore, each point in the torus injected at the boundary has a unique velocity, embodied by a uniform motion. This approach is used in the spheromak model \citep{Verbeke19}. The second strategy models a self-similar evolution at the boundary, as demonstrated in the FRi3D model \citep{Maharana22}. At $0.1$ AU, all intersection points between the torus and the EUHFORIA boundary have only a purely radial speed, equivalent to $v_{r0}$.

The numerical implementation accommodates both methods. However, numerous tests have revealed that simulations exhibit greater stability when using self-similar evolution, as opposed to configurations where each point can have poloidal and toroidal speeds. In the latter scenario, the injected CME is more likely to expand unnaturally upon entering the simulation domain, in turn making the simulation unstable. Therefore, in this work, we consider cases where the injected speed at the boundary is purely radial.

Figure~\ref{fig:speed} shows the impact of the initial radial speed, $v_{r0}$, on the profile obtained in EUHFORIA. The geometrical and thermodynamic properties of the injected tori are identical to the reference cases presented in the previous section (cf.\ Sect.~\ref{sec:reference}) with $C_{\alpha}=1$ for mMT and $\alpha_{S}=1$ for Soloviev models. Only the initial speed varies from $400$~km/s to $1000$~km/s.

As observed in Fig.~\ref{fig:speed}, a higher initial speed results in a higher shock speed at the time of arrival at Earth. A higher speed, i.e.,\ $v_{r0}=1000$ km/s, results in a stronger shock, and thus a slightly more abrupt increase in velocity compared to slower cases. However, the maximum speed reached is in all cases lower than the injected speed, indicating that the magnetic structure slows down during its propagation. This deceleration may be caused by a drag force acting on the CME by the surrounding solar wind (cf.\ Sect.~\ref{sec:B0}).

In Fig.~\ref{fig:speed}, the amplitude of the density, magnetic profiles, and plasma $\beta$ differs according to the initial speed of the CME. The higher the latter, the higher/lower the local minimums/maximums. This result is consistent with the in situ measurement of the ACE spacecraft. Indeed, by comparing a set of fast CMEs and a set of slow CMEs, \citet{Masias16} found that the median value of $|B|$ at Earth is higher in the high-speed group than in the slower group \citep[cf.\ also]{Regnault20}. In EUHFORIA, one of the possible reasons for the difference in amplitudes is that, depending on the initial speed, the Earth does not cross the CME at the same location. In Figs.~\ref{fig:Blon} and \ref{fig:Bclt}, for example, one can see that the longitudinal and colatitudinal fields are not uniformly distributed in the simulation. Thus, a change in the impact position can result in a modification of the amplitude of the different quantities \citep[e.g][]{Lepping90}.

Furthermore, the time profiles in Fig.~\ref{fig:speed} seem to indicate substantial compression of the magnetic structure against the strong shock in the cases where $v_{r0}=1000$ km/s. In EUHFORIA, the torus hole does not have the same speed as the torus itself. When the radial speed is high, the back of the torus arrives faster in the domain and catches up more quickly with the front of the torus. The back part of the torus, having had less time to expand, finds itself compressed between the surrounding solar wind and the back. As a result of this compression, the maximum and minimum local values reached by the $B_{x}$, $B_{y}$, and $B_{z}$ components at the front of the CME are higher when the initial speed is increased.

The sheath, delineated by the first bump of $|B|$ and $B_{y}$ in the mMT CMEs, is also more compressed when $v_{r0}=1000$ km/s due to the fast-moving magnetic ejecta from behind. In contrast, in slow CMEs, the magnetic field and plasma have enough time to approach a near-pressure balance. As a result, an expansion of the sheath can be expected to mirror the trend followed by the CME \citep{Demoulin09}.

The trends of the magnetic field components are also impacted by the change in initial speed (cf.\ Fig.~\ref{fig:speed}). For the $B_{x}$ component, there are some differences between the various cases because the magnetic structure is not traversed at exactly the same position at $1$ AU. The most significant difference is related to the temporal evolution of $B_{y}$. In the fastest cases, i.e.,\ with $v_{r0}=1000$ km/s, we see a positive bump in the mMT CME and a negative one in the Soloviev CME at the end of the magnetic ejecta traverse. These bumps are not observed with the slower CMEs (i.e.,\ with $v_{r0}=400$ km/s and $v_{r0}=600$ km/s). We conclude that these new trends are a trace of the back of the torus, which was characterized by a positive $B_{y}$ in the mMT model and a negative $B_{y}$ in the Soloviev solution (cf.\ Fig.~\ref{fig:theoretical}).

According to the theoretical profiles (cf.\ Fig.~\ref{fig:theoretical}), a change in the sign of the $B_{z}$ component is expected in the back part of the CME. However, this change in sign occurs only at the very end of the CME. Therefore, we suggest that this part of the CME does not cover a sufficiently large region to be clearly perceptible in the $B_{z}$ profile and that it is eroded by the interaction with the solar wind in the wake of the front part of the CME.

To circumvent issues of unnatural magnetic field compression and potential inaccuracies in predictions, CMEs might be better modeled using a half-torus, i.e.\ by taking only the front part of the torus into account. In practical terms, this involves injecting only half of the toroidal CME model by halting its passage across the EUHFORIA boundary when the center of the torus reaches 0.1~AU. This approach, which will be explored in a second paper, is employed by \citet{Singh20} who simulated the 12 July 2012 event by introducing half of a spheromak into a realistic background solar wind. Contrary to a full torus, the injected CME geometry then closely resembles a flux rope featuring a curved front with two legs. However, it is worth noting that in \citet{Singh20}, the CME is initially superimposed onto the solar wind rather than being gradually introduced into the domain via the inner boundary.

\subsection{Impact of the initial magnetic field strength} \label{sec:B0}

\begin{figure}[h!]
    \centering
    \includegraphics[width=0.5\textwidth]{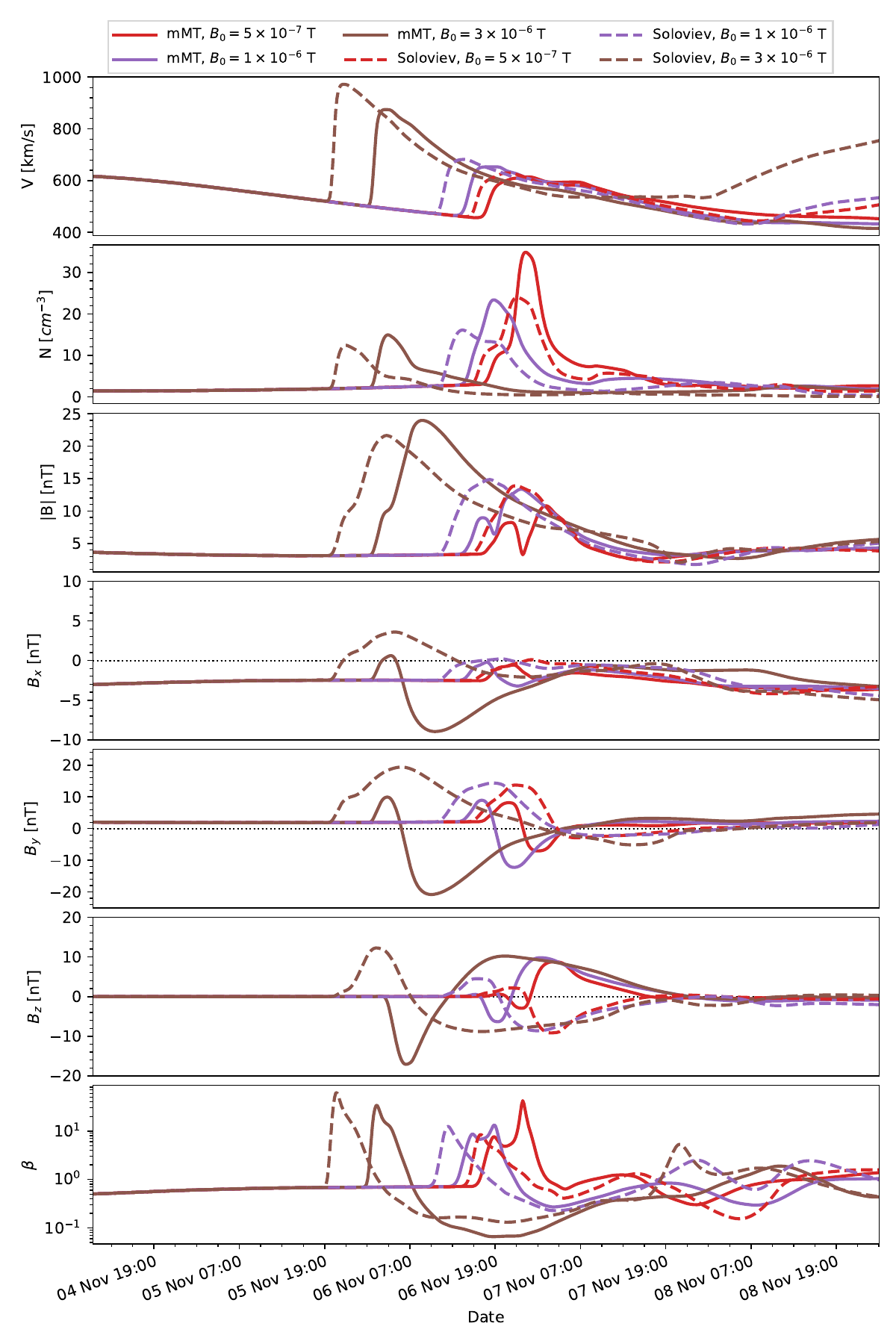}
    \caption{Same as Fig.~\ref{fig:reference}. The only difference between the different CMEs is the initial magnetic field strength $B_{0}$. For CMEs computed from the Turner solution, $B_{0}$ is equal to $5\times 10^{-7}$ T, $1\times 10^{-6}$ T and $3\times 10^{-6}$ T; while it is $1\times 10^{-7}$ T, $2\times 10^{-7}$ T and $6\times 10^{-7}$ T in the Soloviev cases.}
    \label{fig:B}
\end{figure}

To modify the amplitude of the magnetic field profiles, one possibility is to increase the magnetic field $B_{0}$, and consequently the magnetic flux. We used the same geometry as our reference CMEs (cf.\ Sect.~\ref{sec:speed}) and performed simulations with a weaker magnetic flux using a magnetic field that is half as strong as in our reference cases i.e.,\ $B_{0}=5\times 10^{-7}$~T. We also ran simulations with a magnetic field three times larger, i.e.,\ $B_{0}=3\times 10^{-6}$~T.

The first significant difference among the simulations is the computation time. The reference simulations require 23 minutes and 24 seconds for the mMT case and 37 minutes and 39 seconds for the Soloviev one, utilizing the cluster described in Sect.~\ref{sec:setup}. The Soloviev CME, despite having almost the same geometry, magnetic, and kinetic properties, thus requires more computation time when $B_{0}$ equals $1\times 10^{-6}$ T. Since the equations governing the two models are not identical, the time required to calculate them numerically can also vary.

Furthermore, the computation time is intrinsically related to the Courant-Friedrichs-Lewy (CFL) condition, which ensures the stability of the numerical method. Given that the Soloviev CME has slightly more intense temporal profiles than those observed with the mMT CME, it is inferred that more significant gradients exist within the numerical domain. These pronounced gradients can result in swift variations in the domain, increasing the effective propagation speed of disturbances. To maintain stability according to the CFL condition, a rise in the propagation speed leads to a decrease in the time step size. Consequently, a smaller time step requires more steps to cover an equivalent time interval, thus extending the overall computational duration.

The reduction of the value of $B_{0}$ by a factor 2 decreases the computation time to 15 minutes and 19 seconds for the mMT CME and to 13 minutes and 34 seconds for the Soloviev CME. Since the field $B_{0}$ is only a constant used to scale the magnetic field obtained in both CME models, it has no influence on the calculation time of the CME properties during its injection at 0.1~AU. Therefore, if changing the value of $B_{0}$ affects the simulation time, it is because the resolution of the MHD equations requires more or less time, depending on the amplitude of the magnetic field, once the CME has entered the domain.

Indeed, increasing $B_{0}$ requires more computational resources, leading to a considerable increase in computation time. It becomes 27 minutes and 12 seconds for Soloviev and 4 hours and 14 minutes for mMT. In the latter cases, the extended simulation time indicates the solver's difficulty in converging when the gradients between the CME's magnetic field and the solar wind's properties are significantly high at the inner boundary.

Examining the thermodynamic and magnetic profiles in Fig.~\ref{fig:B}, we observe that the larger the magnetic field $B_{0}$, the higher the speed. \citet{Pal2018}, by studying a set of 30 CMEs, also established a positive correlation between their magnetic field strength and their velocity, as determined from white-light images at $10~\;R_{\odot}$. They found that, statistically, CMEs with large $B_{0}$ expand faster than CMEs with a lower magnetic field strength. This is consistent with our observations. However, their statistical analysis did not definitively determine whether this speed difference is a result of a higher ejection speed influenced by the initial magnetic field strength during the eruption or whether it can also be attributed to physical processes during propagation. In our simulations, the latter scenario is illustrated, as our CMEs are all inserted at $21.5~\;R_{\odot}$ with the same initial velocity.

Finally, the speed at Earth depends not only on the initial speed (as discussed in Sect.~\ref{sec:speed}), but also on the magnetic properties of the model. This phenomenon has also been observed in other numerical simulations. For example, by studying the propagation of different flux ropes in the corona, \citet{Linan23} discovered that the maximum speed of the CME at 0.1~AU is directly proportional to the initial magnetic field strength \citep[see also][]{Regnault23}.

The influence of the magnetic field on the propagation speed stems from the balance of forces acting on the CME, which can be described by the following equation:

\begin{eqnarray}
F&=&m\frac{d^{2}R}{dt^{2}} \\
&=&F_{Lorentz}+F_{drag},
\end{eqnarray}

where F is the total force, $m$ the CME mass, $t$ the time, and $R$ is the position of the center of mass. The expanding CME experiences two main forces, the first being the Lorentz force $F_{Lorentz}$ \citep[e.g.,][]{Chen10,Sachdeva15}, which is proportional to $\mathbf{J} \times \mathbf{B}$ where $\mathbf{J}$ is the current density. This force allows for the radial expansion of the CME. The second,$F_{drag}$ is the drag force that tends to restrain the CME, explaining why it gradually slows down during propagation \citep{Chen10,Subramanian12,Sachdeva15,Sachdeva17}. Increasing the magnetic field $B_{0}$ alters the balance between these two forces by strengthening the Lorentz force, resulting in the acceleration of the CME. It is also important to note that the expansion of the CME depends not only on its initial velocity and magnetic properties but also on the characteristics of the background solar wind, particularly through the drag force. For instance, by examining a sample of more than a hundred ICMEs, \citet{Vrvsnak07} suggested a linear least squares fit between the Sun-Earth transit time of interplanetary coronal mass ejections and both their speeds and the solar wind velocity.

Increasing the initial magnetic field $B_{0}$ (cf.\ Fig.~\ref{fig:B}) logically increases the maximum amplitude reached by the total magnetic field $|\mathbf{B}|$. However, in these cases, the maximum magnetic field is weaker than the initial magnetic field $B_{0}$. This means that the magnetic field carried by the CME decreases during its propagation through the heliosphere \citep{Leitner07, Liu05, Winslow15}.

The increase in $B_{0}$ also modifies the trend of $|B|$ (cf.\ Fig.~\ref{fig:B}). For the CMEs computed from the mMT solution, we see two distinct local maxima in the cases where $B_{0}=5\times10^{-7}$ T and $B_{0}=1\times10^{-6}$ T, while there is just a slight change in the slope of the magnetic field increase when $B_{0}=3\times10^{-6}$ T. This change is related to the transition between the sheath and the magnetic ejecta (cf.\ Sect.~\ref{sec:reference}). As explained in Sect.~\ref{sec:speed}, increasing the speed leads to compression of the sheath. This compression is also visible in the profile of plasma $\beta$, which is broader in the cases where the initial magnetic field is the lowest.

In both models, the trends of the components $B_{y}$, $B_{z}$ are not significantly impacted by changes in the initial magnetic field (cf.\ Fig.~\ref{fig:B}). Only the amplitude and width of the profiles are modified. In the extreme case where $B_{0}=3\times10^{-6}$ T in the mMT model, the $B_{x}$ component follows the same trend as the $B_{y}$ component, which was not discernible with a weaker initial magnetic field. As the speed is higher than in other cases, the Earth does not cross at the front of the CME, resulting in a significant $B_{x}$ component as observed in the theoretical profiles (cf.\ Fig.~\ref{fig:theoretical}). However, at each moment, the value of the $B_{x}$ field remains lower than those of the components $B_{y}$ and $B_{z}$.

It is also worth noting that even though the CMEs with a magnetic field strength of $B_{0}$ exhibit speeds close to those obtained in the previous section, where the CMEs were inserted with an initial speed of $v_{r0}=1000$ km/s (cf.\ Fig.~\ref{fig:speed}), the various profiles do not display the same characteristics. Especially, all the profiles with $B_{0}=3\times10^{-6}$ T are broad, in contrast to the compression of profiles observed when $v_{r0}=1000$ km/s in Fig.~\ref{fig:speed}. Furthermore, in Sect.~\ref{sec:speed}, the presence of a bump in the $B_{y}$ component is mentioned, indicating the traversal of the back part of the torus when $v_{r0}=1000$ km/s. We suggest that these features are not observed when $B_{0}=3\times10^{-6}$, since the insertion of the torus still occurs at the speed of $v_{r0}=1000$ km/s. The acceleration of the CME happens gradually during its propagation. Additionally, due to the high magnetic field and the predominance of the Lorentz forces, the front part of the torus is more free to expand spatially and is no longer compressed between the shock and the back part.
\begin{figure}[ht!]
    \centering
    \includegraphics[width=0.5\textwidth]{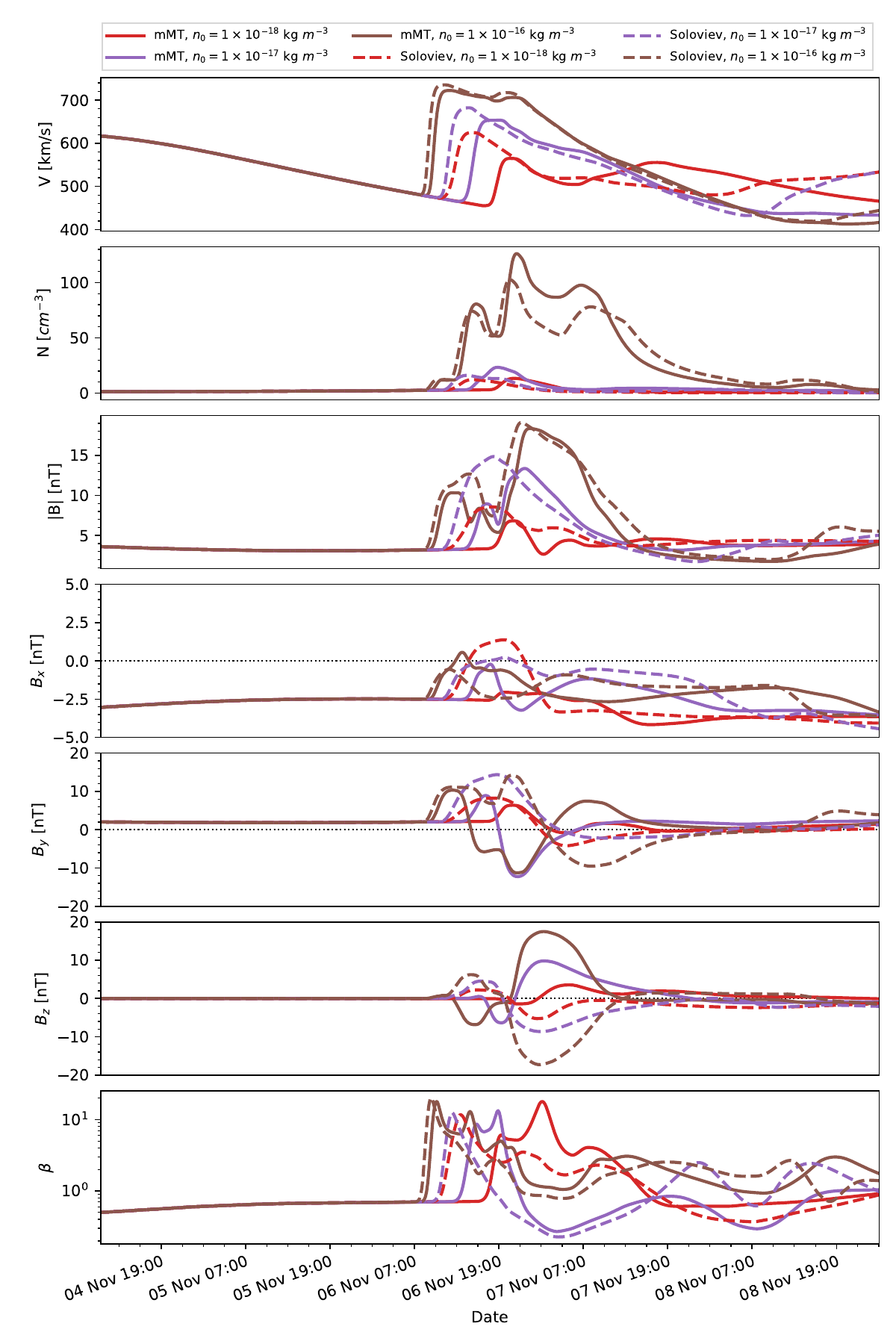}
    \caption{Same as Fig.~\ref{fig:reference}. The only difference between the different cases computed from the same CME model is the initial mass density, $n_{0}$. For both CME solution, the initial mass density is $n_{0}=1\times 10^{-18}\;$kg\,m$^{-3}$, $n_{0}=1\times 10^{-17}\;$kg\,m$^{-3}$ and $n_{0}=1\times 10^{-16}\;$kg\,m$^{-3}$.}
    \label{fig:density}
\end{figure}
Finally, in Fig.~\ref{fig:B}, it is also seen that in both models, increasing the initial field leads to a decrease in the maximum density measured at the Earth's level. This observation will be explained in the following section (cf.\ Sect.~\ref{sec:density}).

\subsection{Impact of the initial mass density} \label{sec:density}

\begin{figure*}[ht!]
    \begin{subfigure}{0.49\linewidth} 
    \centering\includegraphics[width=\linewidth]{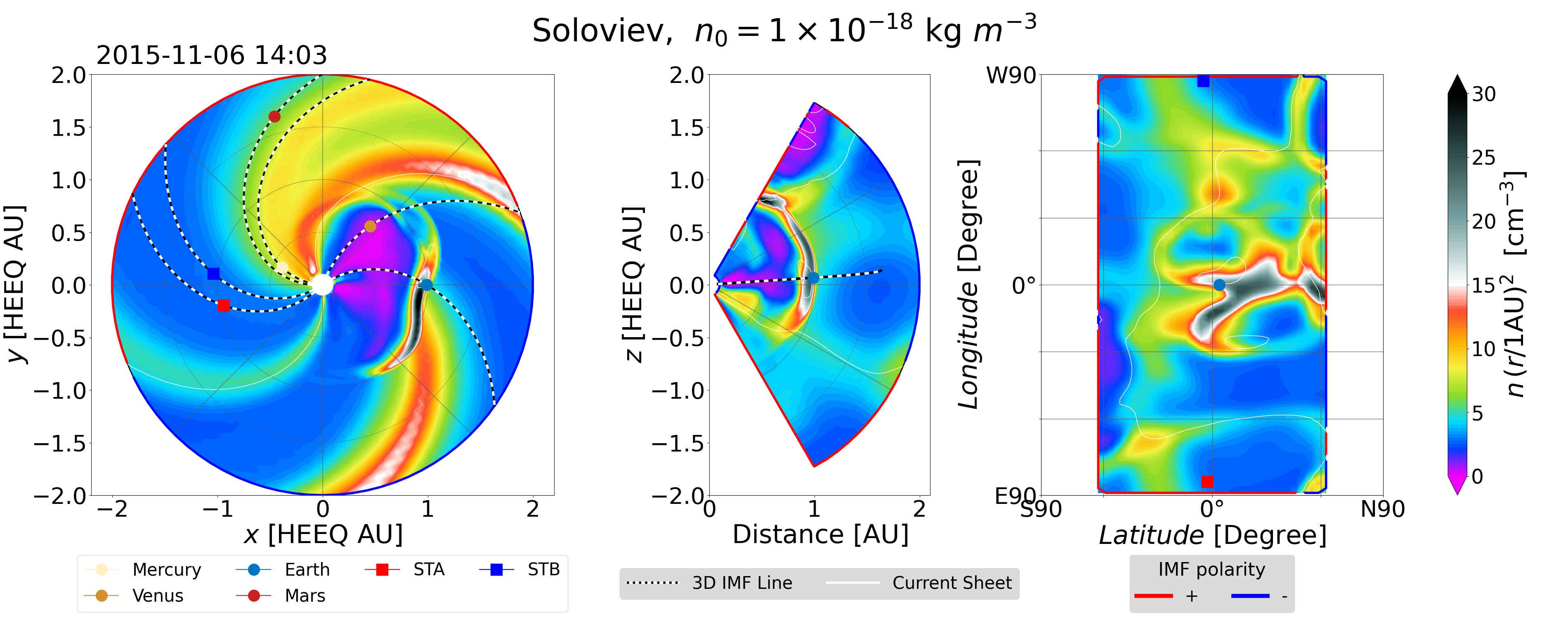}
    \caption{Soloviev CME with $n_{0}=1\times 10^{-18}\;$kg\,m$^{-3}$}\label{fig:na}
    \end{subfigure}
    \hfill
    \begin{subfigure}{0.49\linewidth}
    \centering\includegraphics[width=\linewidth]{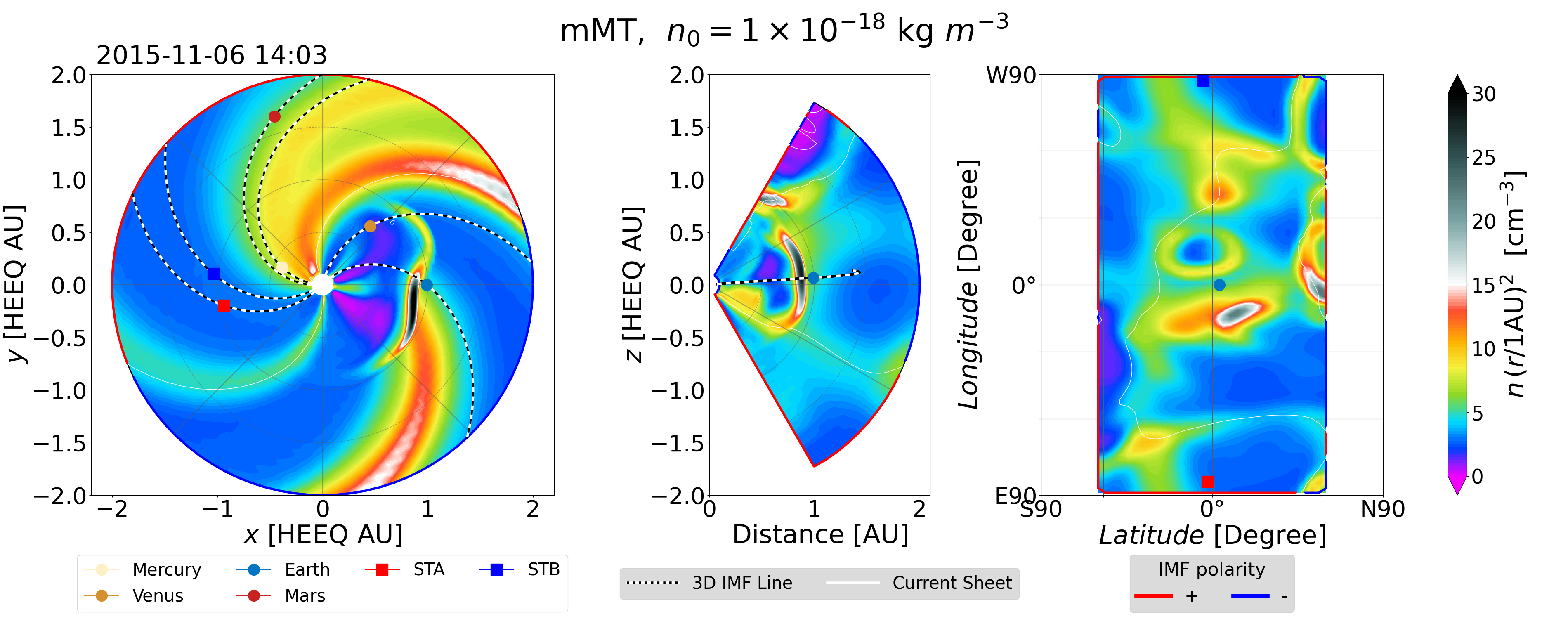}
    \caption{mMT CME with $n_{0}=1\times 10^{-18}\;$kg\,m$^{-3}$}\label{fig:nb}
    \end{subfigure}
    \par\bigskip
    \begin{subfigure}{0.49\linewidth} 
    \centering\includegraphics[width=\linewidth]{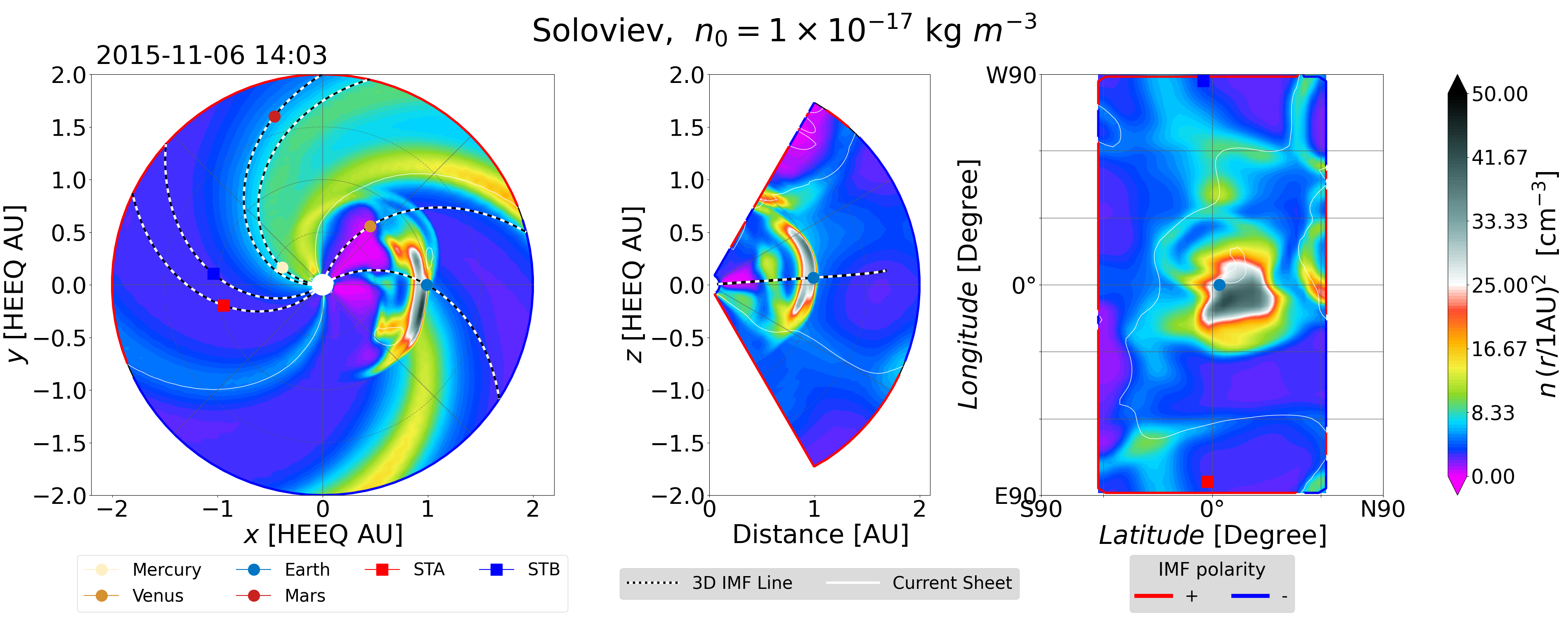}
    \caption{Soloviev CME with $n_{0}=1\times 10^{-17}\;$kg\,m$^{-3}$}\label{fig:nc}
    \end{subfigure}
    \hfill
    \begin{subfigure}{0.49\linewidth}
    \centering\includegraphics[width=\linewidth]{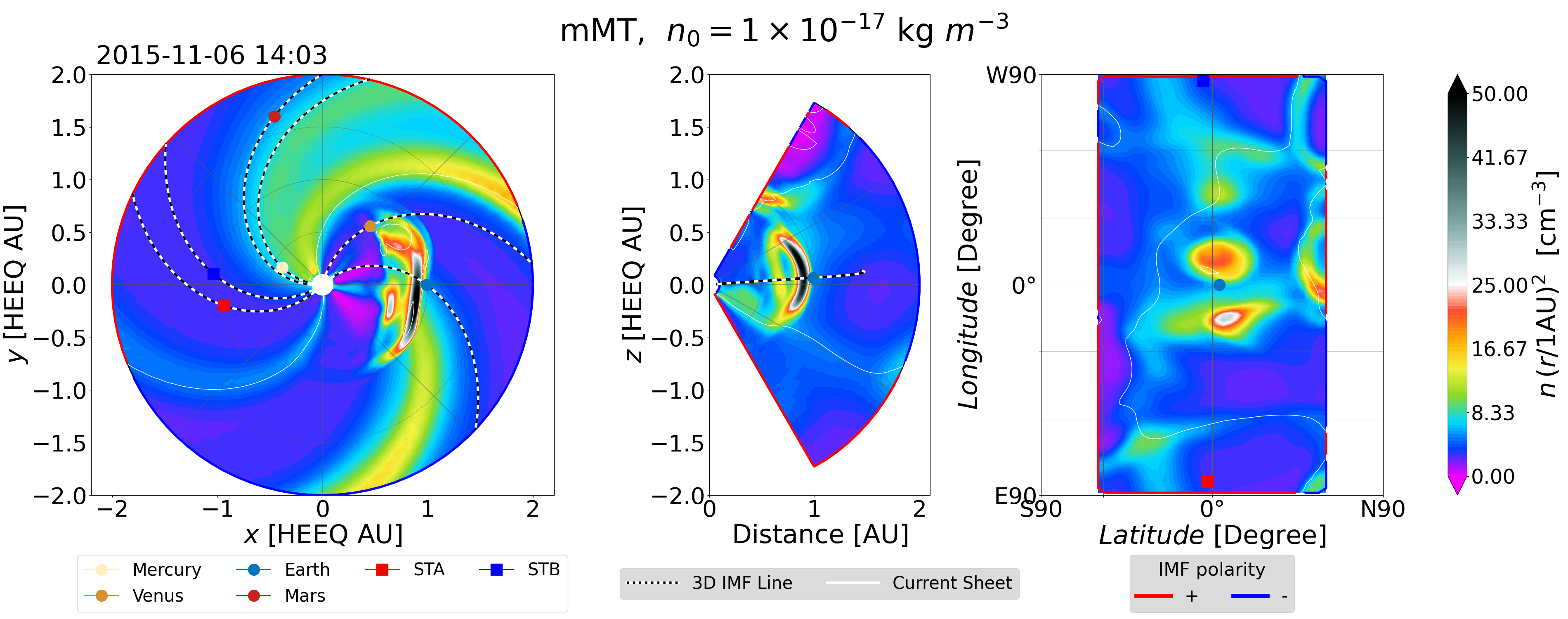}
    \caption{mMT CME with $n_{0}=1\times 10^{-17}\;$kg\,m$^{-3}$}\label{fig:nd}
    \end{subfigure}
    \par\bigskip
    \begin{subfigure}{0.49\linewidth} 
    \centering\includegraphics[width=\linewidth]{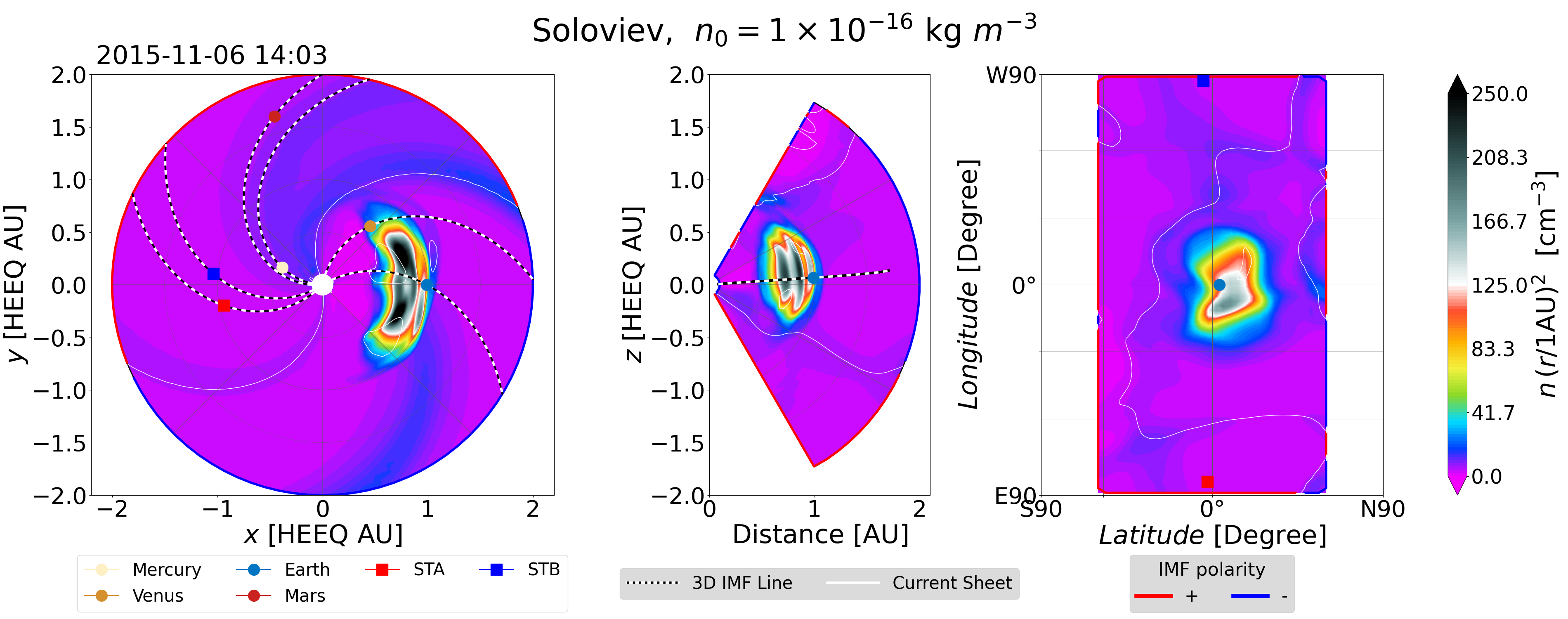}
    \caption{Soloviev CME with $n_{0}=1\times 10^{-16}\;$kg\,m$^{-3}$}\label{fig:ne}
    \end{subfigure}
    \hfill
    \begin{subfigure}{0.49\linewidth}
    \centering\includegraphics[width=\linewidth]{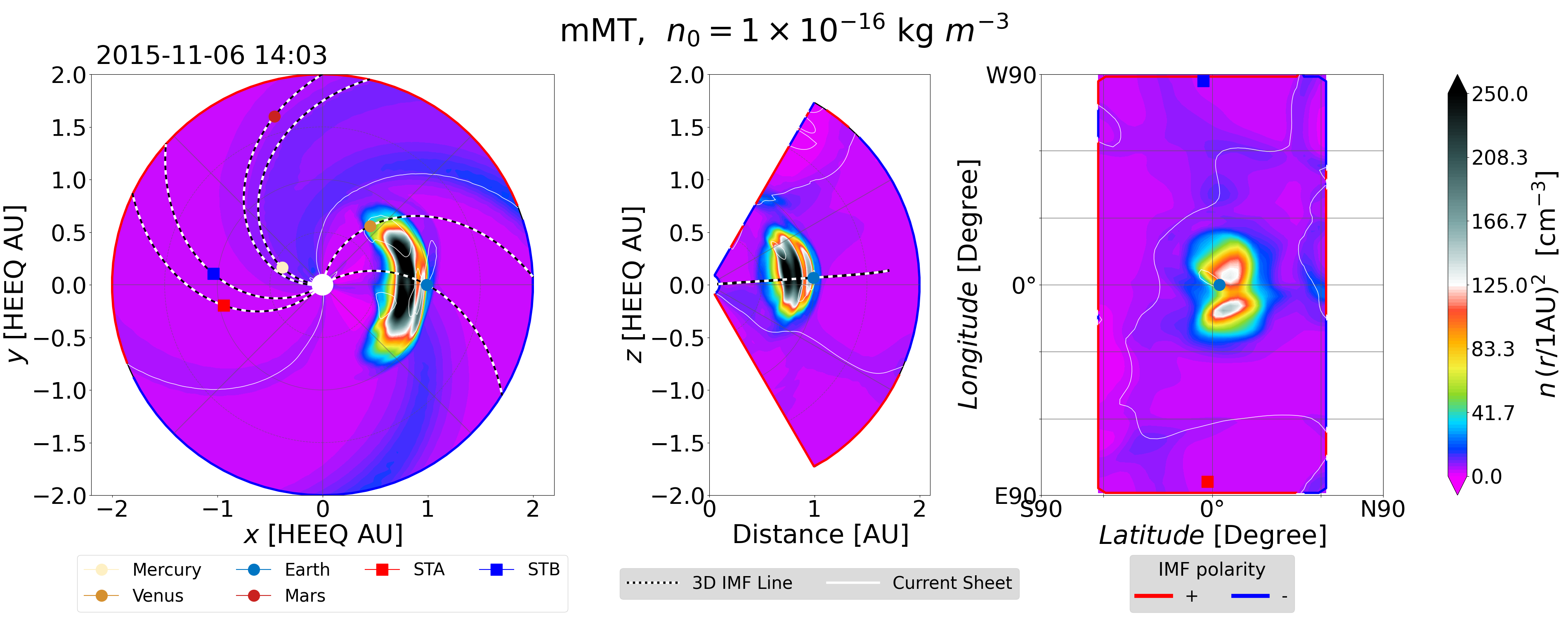}
    \caption{mMT CME with $n_{0}=1\times 10^{-16}\;$kg\,m$^{-3}$}\label{fig:nf}
    \end{subfigure}
    \caption{Distribution of scaled mass density at the moment of arrival at Earth as a function of the initial CME density, $n_{0}$, and the CME model used. The right panels show simulations with the mMT CME model, while the left panels show simulations with the Soloviev CME model. Three different initial mass densities were used, from top to bottom: $n_{0}=1\times 10^{-18}\;$kg\,m$^{-3}$, $n_{0}=1\times 10^{-17}\;$kg\,m$^{-3}$ and $n_{0}=1\times 10^{-16}\;$kg\,m$^{-3}$. The different legends related to the EUHFORIA vizualisation tool are described in Fig.~\ref{fig:Blon}.} \label{fig:dens}
\end{figure*}

The density and temperature of the solar wind at EUHFORIA's inner boundary were adjusted to model a CME's transit. The newly determined density and temperature are uniform at all intersection points, in line with the CME models already implemented in EUHFORIA \citep[e.g.,][]{Scolini19}.

Under the ideal MHD paradigm, upon which the solver is based, imposing a constant density and temperature results in a constant pressure within the torus. However, to achieve the Soloviev equilibrium derived from the Grad-Shafranov equation, the pressure profile is assumed to be linear (cf.\ Sect.~\ref{sec:Soloviev_theory}). Despite this, we chose to adopt constant density and temperature to facilitate numerical computation and comparison between the two CME models employed in our study. Moreover, even if we insert a CME in perfect equilibrium, the equilibrium will be immediately broken upon insertion anyway because the external environment changes drastically.

Figure \ref{fig:density} shows the different thermodynamic and magnetic profiles obtained in EUHFORIA using three different initial mass density : $n_{0}=1\times 10^{-18}\;$kg\,m$^{-3}$, $n_{0}=1\times 10^{-17}\;$kg\,m$^{-3}$, and $n_{0}=1\times 10^{-16}\;$kg\,m$^{-3}$. In this figure, it is apparent that all profiles are affected by the change in density.

As mentioned earlier (cf.\ Sect.~\ref{sec:B0}), the total magnetic field of the CME decreases during its propagation. Similarly, due to the expansion of the CME, the density also decreases \citep{Lugaz05}. Furthermore, when the density is high, specifically $n_{0}=1\times 10^{-16}\;$kg\,m$^{-3}$, the inertia of the plasma inside the CME counteracts both magnetic expansion and dissipation. At the moment of impact with Earth (cf.\ Figs.~\ref{fig:ne} and \ref{fig:nf}), the CME is characterized by a global large, compact magnetic structure with high density and magnetic field. In Figs.~\ref{fig:ne} and \ref{fig:nf}, in the meridional cross-section, we can notably distinguish several regions of high intensity corresponding to the sheath, the front part of the torus, the torus hole, and the rear of the torus. These different regions have "pancake" shapes, resulting from the pressure gradients between the CME and the ambient solar wind \citep{Riley04}. Moreover, we can add that since the magnetic field is high within the torus, so is the Lorentz force, leading to a significant velocity (cf.\ Sect.~\ref{sec:speed}). 

Conversely, when the density is lower, i.e.,\ $n_{0}=1\times 10^{-18}\;$kg\,m$^{-3}$ (cf.\ Figs.~\ref{fig:na}, \ref{fig:nb}), the CME undergoes over-expansion after its insertion into the domain, leading to a significant decrease in both density and magnetic field. As a result, as illustrated in Figure~\ref{fig:density}, the amplitude of the magnetic field, density, and velocity profiles is higher in CMEs with a density of $1\times 10^{-18}\;$kg\,m$^{-3}$ compared to CMEs with lower initial mass densities. These trends are also detailed in \citet{Maharana22}, who obtain similar results by varying the initial density of the FRi3D CME model in EUHFORIA.

Altering the initial density of the CME also leads to a change in the distribution of the longitudinal magnetic field (cf.\ Figs.~\ref{fig:bcltf} and \ref{fig:bclth}) and the colatitudinal magnetic field (cf.\ Figs.~\ref{fig:bcltf} and \ref{fig:bclth}). This results in a modification of the trends of $B_{y}$ and $B_{z}$, as shown in Fig.~\ref{fig:density}. Specifically, a prolonged phase where $B_{y}$ is positive in the mMT profile and negative in the Soloviev profile is observed towards the end of the magnetic ejecta when the initial density is high, i.e.,\ $n_{0}=1\times 10^{-16}\;$kg\,m$^{-3}$. As mentioned in Sect.~\ref{sec:speed}, and according to the theoretical profile (cf.\ Fig.~\ref{fig:theoretical}), this corresponds to the passage through the rear of the torus.

In simulations with the CMEs having the highest initial densities, i.e.,\ $n_{0}=1\times 10^{-16}\;$kg\,m$^{-3}$, the shape of the density profile also reflects the clear traversal of different magnetic structures composing the different parts of the torus as shown in the two dips in plasma beta in Fig.~\ref{fig:density} (7 Nov and 8 Nov).

Finally, looking at the density distribution in Fig.~\ref{fig:dens}, it is also worth noting that in all the simulations, the maximum density is located at the nose of the CME. Therefore, if Earth were to traverse the flank of the CME, the density encountered would be lower than that if it had passed through the front part. Since the impact position depends on the speed of the CME, the amplitude of the density profile obtained varies according to the initial speed and magnetic field imposed (cf.\ Figs.~\ref{fig:speed} and \ref{fig:B}).

\subsection{Impact of major and minor radii} \label{sec:impacta}
The torus must begin its traversal of the boundary just after the insertion time. In other words, before the insertion time, there are no intersection points between the CME and the inner boundary. To ensure this, the following equation must be satisfied:
\begin{equation}
t_d+a+R_{0}=21.5~\;R_{\odot}. \label{eq:limit}
\end{equation}
Therefore, the size of the CME is limited as its maximum value, $a+R_{0}=21.5~\;R_{\odot}$.

\begin{figure}[h]
    \centering
    \includegraphics[width=0.5\textwidth]{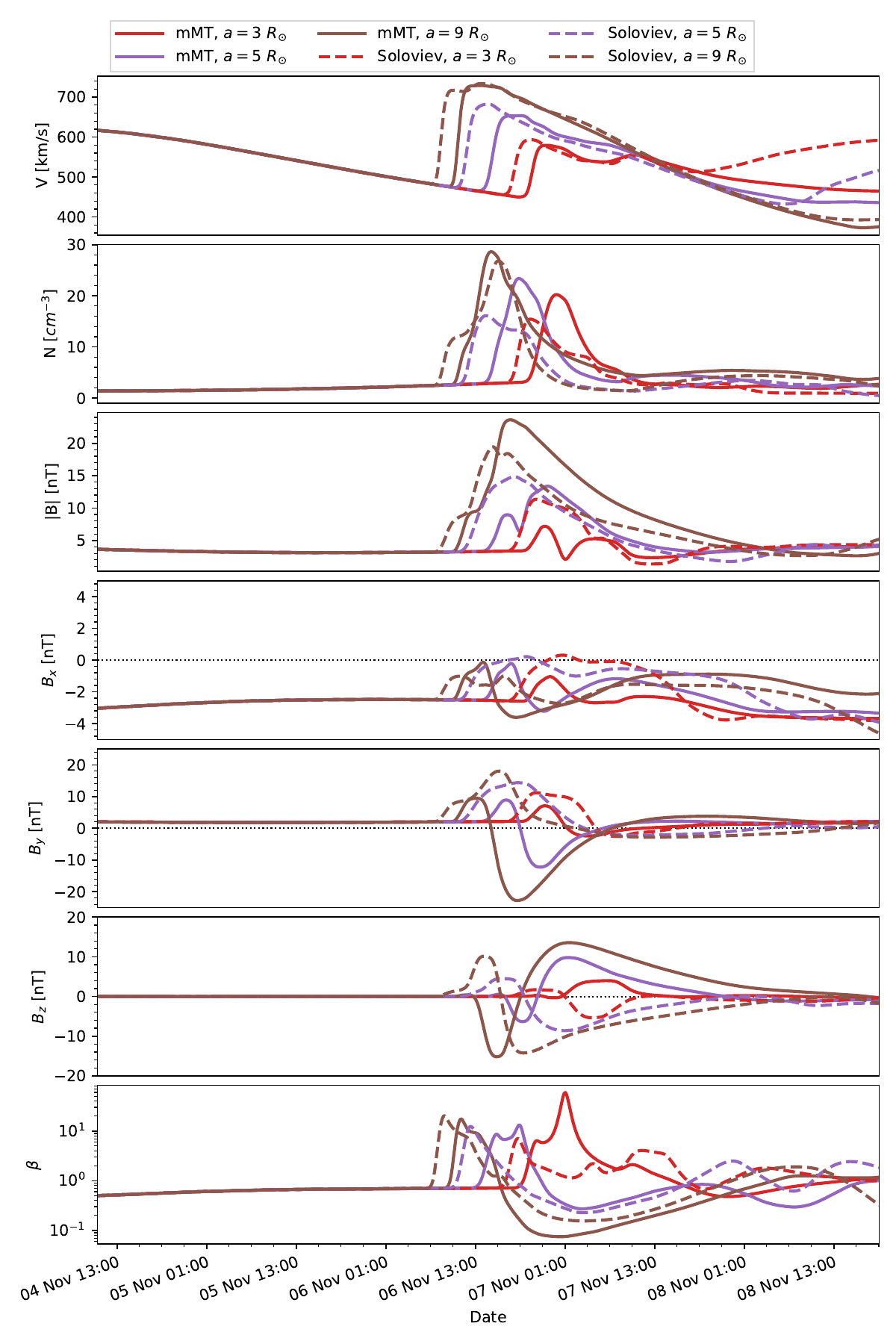}
    \caption{Same as Fig.~\ref{fig:a}. The only difference between the different CMEs is the minor radius, $a=3\;R_{\odot}$, $a=5\;R_{\odot}$ and $a=9\;R_{\odot}$. The major radius is fixed to $R_{0}=10\;R_{\odot}$}
    \label{fig:a}
\end{figure}

Figure~\ref{fig:a} shows the time evolution profiles obtained in EUHFORIA for CMEs with varying minor radii $a$. In these instances, the major radius is fixed at $R_{0}=10\;R_{\odot}$. Consequently, by modifying the minor radius, we have also adjusted the initial distance from the center of the torus to the center of the domain, i.e.,\ $t_d$, according to Eq. \ref{eq:limit}.

As the minor radius increases, so does the spatial size of the CME. Therefore, the larger $a$, the wider the profiles obtained in Fig.~\ref{fig:a} will be. On the other hand, increasing the minor radius without altering the initial magnetic field $B_{0}$ results in a rise in magnetic flux. Therefore, the magnetic profiles for the case of $a=9\;R_{\odot}$ are comparable to those obtained when $B_{0}=3\times 10^{-6}\;$T in Sect.~\ref{sec:B0}. As previously described, the amplitude of the various magnetic profiles and the speed of the CME increase with the minor radius. However, the computation time also increases. It goes from 23 minutes and 24 seconds for the reference mMT CME with $a=5\;R_{\odot}$ to 1 hour and 11 minutes for the mMT CME with $a=9\;R_{\odot}$. The simulation using a Soloviev model with $a=9\;R_{\odot}$ requires approximately three times the computation time as the reference simulation with $a=5\;R_{\odot}$.

All different CMEs shown in Fig.~\ref{fig:a} have the same initial density $n_{0}$. As the minor radius increases, so does the volume of the torus and, hence, the total mass of the CME injected across the boundary. Therefore, the maximum of the obtained density profile is higher for $a=9\;R_{\odot}$ than for $a=3\;R_{\odot}$, as described in Sect.~\ref{sec:density}. However, it should be recalled that the density profile's amplitude depends on the impact position between the CME and Earth.

Finally, altering the torus geometry, for example, by adjusting the minor radius, changes the magnetic and thermodynamic properties of the CME and thus the resulting profiles, as described in previous sections. It is important to note that, numerically, the solver (i.e.,\ EUHFORIA) allows the use of a minor radius close to the major radius, i.e.,\ $R_{0}/a\approx1$. Theoretically, the magnetic field is force-free only for large aspect ratios ($R_{0}/a\gg1$). Therefore, we can conclude that the force-free assumption is not a limiting criterion and that it is not necessary in EUHFORIA to restrict the range of minor and major radii to satisfy $R_{0}/a\gg1$.

\section{Conclusion} \label{sec:conclusion}

In order to introduce new CME models in EUHFORIA to fill the gaps in the models currently used, we implemented two magnetic configurations with a toroidal geometry. The first CME model is derived from the modified Miller-Turner (mMT) model (cf.\ Sect.~\ref{sec:Millerturner}). The second corresponds to the Soloviev equilibrium, an analytical solution to the Grad-Shafranov equation (cf.\ Sect.~\ref{sec:Soloviev_theory}). Although both models possess a toroidal geometry, the Soloviev CME model offers more free parameters. Indeed, unlike the mMT model, where the twist is fixed, it is possible to modify the magnetic helicity of the Soloviev solution. On the other hand, the mMT model has a circular cross section, while the poloidal aspect can be modified in the Soloviev CME model by adjusting the triangularity and the elongation (cf.\ Sect.~\ref{sec:geometrysoloviev}). The distribution of the magnetic field within the torus is also different in both models, which justifies the implementation of two configurations (cf.\ Sect.~\ref{sec:profileth}).

After presenting these differences analytically, we show the numerical implementation in EUHFORIA (cf.\ Sect.~\ref{sec:implementation}). This implementation is identical in both models. The torus crosses the inner boundary of EUHFORIA with an initially fixed radial speed. At each intersection point between the torus and the boundary, the speed, magnetic field, density, and temperature of the solar wind are modified to model the passage of the CME.

To illustrate the numerical implementation, we performed a series of simulations in which the CME is injected into a realistic solar wind background derived from the GONG magnetogram obtained at 01:04 UT on November 4, 2015. We examined the impact of different initial parameters on the magnetic and thermodynamic profiles at Earth in EUHFORIA. From these profiles, our main findings were as follows:

\begin{itemize}
\item Both models are stable and can be used in EUHFORIA. The magnetic profiles obtained at Earth are consistent with the distribution of the magnetic field injected across the boundary.
\item Once within the domain, the CME radially expands with a speed dependent on the radial velocity of the torus center during torus insertion. When the initial velocity is high (approximately $1000\;$km/s), we observed a compression of the magnetic structure between the back of the torus and the shock. This leads to narrow profiles. Examining the distribution of the magnetic field, we noticed the development of a sheath ahead of the mMT CME, the thickness of which also depends on the speed of the magnetic ejecta.
\item Increasing the magnetic field leads to an increase in the Lorentz force, and therefore increasing the speed of the CME. This also allows us to obtain broad profiles with high amplitudes. However, the computation time is directly related to the intensity of the magnetic field.
\item In both models, the CME expansion is directly related to the initial mass density of the torus. When the density is too low (here $n_{0}=1\times 10^{-18}\;$kg\,m$^{-3}$), the CME experiences over-expansion, leading to dispersion of both density and magnetic field. In contrast, when the density is high (e.g., $n_{0}=1\times 10^{-16}\;$kg\,m$^{-3}$), the torus maintains a compact structure. It is then possible to discern the influence of the back of the torus on the magnetic profiles.
\item The amplitude of the local peaks and the duration of the magnetic disturbances can be modified by adjusting the geometry of the torus. The greater the minor radius, the greater the injected magnetic flux and total mass. Therefore, the amplitude of the magnetic field and density can be altered without changing the initial magnetic field strength and the initial mass density.
\end{itemize}

Finally, EUHFORIA enables the use of these two new toroidal models. Despite their relatively simple analytical formulation, it is possible to achieve complex profiles, depending not only on the initial parameters of the CME but also on the impact position with Earth.

Now, the next step is to use these two new toroidal CME models to predict the geoeffectiveness of real CME events. Similar to \citet{Scolini19}, the properties of the torus, such as its size, magnetic flux, and speed, must be deduced from various observations to ensure the most realistic simulation possible, thus providing the most accurate predictions \citep[cf.\ also][]{Maharana22}. This kind of study will allow us to conclude which model is most suitable for modeling real events. In fact, our study alone cannot determine if one model is superior to the other, since both models provide different magnetic profiles and react almost similarly to changes in the initial parameters. However, we suggest that both the mMT and the Soloviev models can provide profiles that closely align with observations, especially since both models exhibit a change in the $B_{z}$ profile sign, as can be measured by an in-situ satellite at 1~AU \citep[e.g.,][]{Regnault20}.

Moreover, by successively launching two spheromak CMEs into the heliospheric models EUHFORIA and PLUTO, respectively, \citet{Scolini20} and \citet{Koehn22} found that the interaction of two moderate CMEs can result in an intense geomagnetic storm. Therefore, it would be insightful to study within EUHFORIA how the thermodynamic and magnetic profiles at Earth are affected by the interaction of multiple CMEs modeled by our mMT and/or Soloviev models.

It may also be interesting to compare the performance of these two models with previously implemented models such as spheromak \citep{Verbeke19} and FRi3D \citep{Maharana22}. Regarding the geometry, given that coronal mass ejections are often approximated as locally cylindrical objects with circular cross sections \citep{Savani11}, the morphology of these two new toroidal models is likely to be closer to observations than the spheromak model, which is spherical. Moreover, the geometry of our toroidal models can be further refined using the radial stretching approach proposed by \citet{Gibson98}. This enhancement would make them even more closely resembling a realistic flux rope. Finally, we note that the computation time required for the two toroidal CMEs (on the order of a few minutes for our reference runs) is closer to that of the spheromak model than that of FRi3D. Indeed, the latter model, with standard initial parameters and a numerical configuration similar to ours, requires between 7 and 10 hours for the calculation, while the spheromak requires between several tens of minutes to a couple of hours \citep{Maharana22}.

We have found, however, that increasing the magnetic flux can cause a massive increase in computation cost because more time steps are needed to capture the time-dependent evolution on the grid. To speed up the computation, it could be worthwhile to implement these two CME models in the new inner heliospheric simulation ICARUS \citep{Verbeke22, Baratashvili22}. Unlike EUHFORIA, ICARUS is based on MPI-AMRVAC \citep{Keppens03,Keppens12,Keppens2023,Xia18}, which uses advanced numerical techniques and radial grid stretching together with adaptive mesh refinement. \citet{Baratashvili22} found that for the same spheromak CME evolving in the same solar wind, ICARUS is up to 17 times faster than EUHFORIA. Furthermore, the adaptive mesh refinement in MPI-AMRVAC could allow the capture of finer details in the simulation that may be missed when using a uniform grid. We plan to validate the implementation of these two new models using observation data in a second paper.

To conclude, the current implementation of the toroidal CMEs in EUHFORIA has laid the groundwork for additional toroidal CME modeling. By addressing all the intricacies specific to a toroidal CME structure, such as determining the insertion points, we have paved the way for a simplified adaptation process. This foundation makes it significantly easier to incorporate other toroidal CME models, such as the Tsuji solution \citep{Tsuji91} or the modified Titov-Démoulin CME model \citep{Titov99,Titov14}. These can now be implemented by simply adjusting the magnetic field within the torus.

\begin{acknowledgements}
The authors thank the anonymous referee for their insightful comments. We acknowledge support from the European Union's Horizon 2020 research and innovation program under the grant agreement N$^o$ 870405 (EUHFORIA~2.0) and from the projects C14/19/089 (C1 project Internal Funds KU Leuven), G.0B58.23N and G.0025.23N (FWO-Vlaanderen), 4000134474 (SIDC Data Exploitation, ESA Prodex-12), and Belspo project B2/191/P1/SWiM. RK acknowledges support by the C1 project TRACESpace funded by KU Leuven and by the European Research Council (ERC) under the European Unions Horizon 2020 research and innovation program (grant agreement No.\ 833251 PROMINENT ERC-ADG 2018) and a FWO project G0B4521N.
\end{acknowledgements}

%
%


\bibliographystyle{aa} 
\bibliography{biblio.bib}

\begin{appendix}
\section{Magnetic fields in the global coordinates}\label{sec:appendix}
\subsection{Miller-Turner transformation system}\label{sec:appendixMT}

According to Sect.~\ref{sec:Millerturner}, the magnetic field for the modified Miller-Turner CME within the torus is defined in the local $(\rho_{l},\phi_{l}, \theta_{l})$ curved cylindrical coordinates. In order to implement in EUHFORIA, we first derive the magnetic fields in the local spherical system ($e_{\rho},e_{\theta},e_{\phi}$) where $\rho$ is the radius, $\theta$ is the colatitude, and $\phi$ is the longitude. The two set of coordinates are related to each other by the relations :
\begin{eqnarray}
    \rho_{l}&=&\sqrt{\rho^{2}+R_{0}^{2}-2\rho  R_{0}\sin\theta} \label{eq:rhol}\\
     \theta_{l}&=&
\begin{cases}
    \arcsin{\frac{\rho\cos\theta}{\rho}},& \text{if } \rho\sin\theta-R_{0}\geq 0\\
    \pi-\arcsin{\frac{\rho\cos\theta}{\rho}},              & \text{otherwise}.
\end{cases}
\end{eqnarray}
The magnetic field in the local spherical system can be written as : 
\begin{eqnarray} 
B_{\rho}&=&\sin(\theta_{l}+\theta) B_{\rho_{l}}+\cos(\theta_{l}+\theta) B_{\theta_{l}} \\
B_{\phi}&=&B_{\phi_{l}} \\
B_{\theta}&=&\cos(\theta_{l}+\theta) B_{\rho_{l}}-sin(\theta_{l}+\theta) B_{\theta_{l}}
\end{eqnarray}
In a similar way, the magnetic field can also be obtained in the local Cartesian system ($e_{Xl},e_{Yl},e_{Zl}$) defined by the coordinates ($X_{l},Y_{l},Z_{l}$) :
\begin{eqnarray} 
B_{Xl}&=&\frac{B_{\rho}X_{l}}{\sqrt{X_{l}^{2}+Y_{l}^{2}+Z_{l}^{2}}}+\frac{B_{\theta}X_{l}Z_{l}}{\sqrt{X_{l}^{2}+Y_{l}^{2}+Z_{l}^{2}}\sqrt{X_{l}^{2}+Y_{l}^{2}}} \nonumber\\&&-\frac{B_{\phi}Y_{l}}{\sqrt{X_{l}^{2}+Y_{l}^{2}}}  \\
B_{Yl}&=&\frac{B_{\rho}Y_{l}}{\sqrt{X_{l}^{2}+Y_{l}^{2}+Z_{l}^{2}}}+\frac{B_{\theta}Y_{l}Z_{l}}{\sqrt{X_{l}^{2}+Y_{l}^{2}+Z_{l}^{2}}\sqrt{X_{l}^{2}+Y_{l}^{2}}} \nonumber\\ &&+\frac{B_{\phi}X_{l}}{\sqrt{X_{l}^{2}+Y_{l}^{2}}} \\
B_{Zl}&=&\frac{B_{\rho}Z_{l}}{\sqrt{X_{l}^{2}+Y_{l}^{2}+Z_{l}^{2}}}-\frac{B_{\theta}(X_{l}^{2}+Y_{l}^{2})}{\sqrt{X_{l}^{2}+Y_{l}^{2}+Z_{l}^{2}}\sqrt{X_{l}^{2}+Y_{l}^{2}}}
\end{eqnarray}
with 
\begin{eqnarray} 
X_{l}&=&\rho\sin\theta\cos\phi \\
Y_{l}&=&\rho\sin\theta\sin\phi \\
Z_{l}&=&\rho\cos\theta 
\end{eqnarray}
It is worth noting that we use the magnetic field in the spherical coordinate system as an intermediate step rather than directly transforming the magnetic field from the local cylindrical coordinate system to the Cartesian coordinate system.

In order to implement the magnetic field in EUHFORIA, we need to switch from the local Cartesian system ($e_{Xl},e_{Yl},e_{Zl}$) to the global Cartesian system ($e_{X},e_{Y},e_{Z}$) of the simulation. The center of the torus in the EUHFORIA system is the point $O_{l}$ at $\rho=t_{d}$, $\theta=\theta_{l}$, and $\phi=\phi_{l}$. The original coordinate system is first translated such as : 
\begin{eqnarray} 
x_{tr}&=&x-t_{d}\sin\theta_{l}\cos\phi_{l} \\
y_{tr}&=&y-t_{d}\sin\theta_{l}\sin\phi_{l} \\
z_{tr}&=&z-t_{d}\cos\theta_{l}, 
\end{eqnarray}
where $(x_{tr}, y_{tr}, z_{tr})$ define the translated system. Then to obtain the local Cartesian system, the new coordinate system is rotated, so that the new X-axis corresponds to the direction of propagation $\boldsymbol{OO_{l}}$, where O is the center of the EUHFORIA domain (cf.\ Fig.~\ref{fig:system}).

\begin{equation}
\begin{pmatrix}
X  \\
Y  \\
Z
\end{pmatrix}
=
\begin{pmatrix}
\sin\theta_{l} & 0 & \cos\theta_{l} \\
0              & 1 & 0               \\
-\cos\theta_{l} & 0 & \sin\theta_{l}
\end{pmatrix}
\begin{pmatrix}
\cos\phi_{l}      & \sin\phi_{l} & 0 \\
-\sin\phi_{l}      & \cos\phi_{l} & 0               \\
0 & 0 & 1
\end{pmatrix}
\begin{pmatrix}
x_{tr}  \\
y_{tr}  \\
z_{tr}
\end{pmatrix}
\end{equation}
Finally, to apply a tilt defined by an angle $\omega$, the resulting system can be rotated along the new X-axis such as :
\begin{equation}
\begin{pmatrix}
X_{rot}  \\
Y_{rot} \\
Z_{rot}
\end{pmatrix}
=
\begin{pmatrix}
1 & 0 & 0 \\
0              &  \cos\omega & - \sin\omega              \\
0 & \sin\omega &  \cos\omega
\end{pmatrix}
\begin{pmatrix}
X  \\
Y  \\
Z,
\end{pmatrix}
\end{equation}
with ($X_{rot}$, $Y_{rot}$, $Z_{rot}$) define the final coordinate system.

\subsection{Soloviev transformation system}

Similarly to the modified Miller-Turner CME model, the Soloviev solution is obtained in a local coordinate system, which differs from the coordinate system used in EUHFORIA. Therefore, a coordinate transformation is necessary. The Soloviev CME model is defined in a specific cylindrical coordinate system, defined by the unit vectors $e_{R_{l}}$, $e_{\phi_{l}}$ with $R_{l}$ the radius, $Z_{l}$ the vertical axis, and $\phi_{l}$ the longitudinal angle. These cylindrical coordinates can be expressed in terms of the local spherical coordinates $(\rho,\theta,\phi)$:
\begin{eqnarray} 
R_{l}&=&\sqrt{(\rho\sin{\theta}\cos{\phi})^{2}+(\rho\sin{\theta}\sin{\phi})^{2}} \\
Z_{l}&=&\rho\cos{\theta} \\
\phi_{l}&=&-\phi\text{.}
\end{eqnarray}
The three components of the magnetic field in the local spherical coordinate system can be obtained from the magnetic field in the cylindrical coordinate system : 
\begin{eqnarray} 
B_{\rho}&=&\sin{\theta} B_{R_l}+cos{\theta} B_{Z_l} \\
B_{\theta}&=&\cos{\theta} B_{R_l}-sin{\theta} B_{Z_l} \\
B_{\phi}&=&-B_{\phi_{l}} 
\end{eqnarray}

Once the components of the magnetic field are obtained in the local spherical coordinate system, they allow access to the vectors in the local Cartesian coordinate system before undergoing a translation and rotations to obtain the different vectors in the global system used in EUHFORIA. The transformation from the components $B_{\rho}$, $B_{\theta}$, $B_{\phi}$ to the vector in the final EUHFORIA coordinate system is described in Sect.~\ref{sec:appendixMT}.

\end{appendix}


\end{document}